\documentclass[preprint,groupedaddress,amsmath,amssymb,
 aps]{revtex4-2}

\usepackage{graphicx}
\usepackage{dcolumn}
\usepackage{bm}
\usepackage{amsfonts}
\usepackage{float}
\usepackage{ulem}
\usepackage{color}
\usepackage{subcaption}
\usepackage{caption}
\usepackage{natbib}

\newcommand{\br}[1]{\left( #1 \right)}
\newcommand{\sbr}[1]{\left[ #1 \right]}
\newcommand{\me}{\mathrm{e}}
\newcommand{\mdd}{\mathrm{d}}
\newcommand{\mi}{\mathrm{i}}
\newcommand{\Ro}{\textrm{Ro}}

\newcommand{\abs}[1]{\left| #1 \right|}
\newcommand{\pa}[1]{\partial_{#1}}
\newcommand{\Ja}[2]{\mathcal{J}\left(#1,#2\right)}
\newcommand{\Jacd}[2]{\mathcal{J}\left(#1;#2\right)}

\newcommand{\bs}[1]{\boldsymbol{#1}}

\newcommand{\ifourier}[1]{\mathcal{F}^{-1}\{#1\}}

\begin{document}


\title{Energy spectrum of two-dimensional isotropic rapidly rotating turbulence}

\author{Peiyang Li}
\email[]{lipeiyang@pku.edu.cn}
\altaffiliation{}
\author{Jin-Han Xie}
\email[]{jinhanxie@pku.edu.cn}
\affiliation{Department of Mechanics and Engineering Science at College of Engineering, and State Key Laboratory for Turbulence and Complex Systems, Peking University, Beijing 100871, P. R. China}


\date{\today}

\begin{abstract}
	We study a two-dimensional isotropic rotating system and obtain both theoretically and numerically a $K^{-2}$ energy spectrum under the rapidly rotating condition ($Ro\ll 1$), which was initially obtained by \citet{Zeman1994} and \citet{Zhou1995}.
In rotating turbulence, the $K^{-2}$ energy spectrum was proposed under the assumption of isotropy, however, the direction selectivity of rotation breaks isotropy, making this $K^{-2}$ spectrum not easily observable.
To fill the gap between theoretical assumptions and realizability, we study the turbulence of inertial waves in an artificial two-dimensional isotropic rotating turbulence system. 
In the limit of a small Rossby number, we asymptotically derive a nonlinear amplitude equation for inertial waves, which gives the $K^{-2}$ spectrum using a strong turbulence argument.
This scaling is justified by numerical simulations of both the amplitude equation and the original system.

	

\end{abstract}


\maketitle

\section{\label{intro}Introduction}

Rotating turbulence arises in a variety of geophysical and engineering applications, such as atmospheric dynamics, ocean currents, and turbine flows \citep{salmon1998lectures,2019VallisBook,Marshall2008,Mayle1991TheRO}. 
The effect of rotation, as predicted by the Talor-Proudman theorem, leads to two-dimensional (2D) behaviour \citep{Deusebio2014,davidson2015turbulence,gallet_2015}, thus, leads to interesting energy cascade behaviour. 
Energy forward cascade to small scales in three-dimensional (3D) turbulence with the Kolmogorov spectrum $E(K)\sim K^{-5/3}$ and $K$ is the wavenumber, while in 2D energy transfers upscales with $E(K)\sim K^{-5/3}$ and enstrophy cascades to small scales with $E(K)\sim K^{-3}$ \citep{Kraichnan1967}.
Rotating turbulence usually has a dual cascade scenario, where a forward energy cascade occurs at small scales and an inverse energy cascade appears at large scales due to rotation \citep{Cho2016ForwardAI,Sharma2018OnTE,Alam2018RevisitingBS}.
The dual cascade is an important phenomenon in geophysical fluid dynamics, as it helps explain the behaviour of atmospheric and oceanic flows, especially the formation of large-scale coherent structures induced by inverse cascade \citep{smithr_yakhot_1994,Cho2001,Shats2007,Marino2015,van_kan_alexakis_2022,Balwada2022}.

Without considering anisotropy, \citet{Zeman1994} proposed a dual spectrum in rotating turbulence separated by a Zeman wavenumber $K_{Z}=\br{\Omega^3/\epsilon}^{1/2}$, where $\Omega$ is the angular velocity and $\epsilon$ is the energy dissipation rate.
When $K \gg K_{Z}$, the energy spectrum follows Kolmogorov’s theory with scaling $K^{-5/3}$, while when $K \ll K_{Z}$, the energy spectrum follows $E(K)\sim K^{-2}$. 
From the perspective of time scale, \citet{Zhou1995} argued that the time scale for inducing turbulent spectral transfer is captured by the 
eddy turnover without rotation, while when rotation is dominant, it is described by the rotating time scale $1/\Omega$, leading to a $E(K)\sim K^{-2}$ spectrum.
The $K^{-2}$ spectrum of rotating turbulence was also obtained by a simple dimensional analysis of wave turbulence theory \cite{Nazarenko2011wave} with a 3D wave triad interaction under the assumption that $k_{\parallel}\sim k_{\perp}$, where $\parallel$ and $\perp$ refer to the direction parallels and perpendicular to the rotation axis, respectively. 

In the ocean, kinetic-energy spectra show a $K_{\perp}^{-2}$ scaling in the submesoscale range \cite{Callies2013}. \citet{Buhler_Callies_Ferrari_2014} separated the contribution of inertial-gravity waves (IGW) from that of geostrophic motion and showed that the IGWs part follows a $K_{\perp}^{-2}$ spectrum.
In a rapidly rotating, strongly stratified fluid, \citet{kafiabad_savva_vanneste_2019} confirm
the validity of the $K^{-2}$ prediction against Boussinesq solutions with assumption of horizontal isotropy.
\citet{Savva_Kafiabad_Vanneste_2021} derived a linear kinetic equation to describe the scattering of IGWs by geostrophic turbulence, and obtained a $K^{-2}$ energy 
spectrum with a cascade from the scale of the internal wave forcing to smaller scales.
\citet{Galtier2003} applied weak wave turbulence theory for 3D incompressible fluids under rapid rotation and analytically found an anisotropic energy spectrum $E\sim k_{\perp}^{-5/2}k_{\parallel}^{-1/2}$ when the transfer time given by the wave kinetic equation is much larger than the inertial-wave period. 
This condition is not valid in the entire wavenumber space, and the weak wave turbulence theory does not describe the dynamics of 2D modes at the lowest order which decouples from 3D waves.

In this paper, we attempt to study whether the isotropic arguments of \citet{Zeman1994} and \citet{Zhou1995} capture certain key features of rotating turbulence, which is hard due to the naturally introduced anisotropic effect by rotation.
Moreover, their discussions are limited by choosing a single finite time scale $1/\Omega$.
However, in a rotating fluid system, the linear dispersion relation is \cite{Nazarenko2011wave}
\begin{equation}
	\omega=\pm \sqrt{\frac{(2\Omega)^2 k_{\perp}^2}{k_{\parallel}^2+k_{\perp}^2}},
\end{equation}
where the frequency of the inertial wave, $\omega$, ranges between $0$ and $2\Omega$, leading to a wide range of time scale.
To be consistent with the isotropy assumptions by Zeman and Zhou, we introduce an isotropic rotating turbulence system, where the linear inertial waves have only a single timescale. 

An effect analogous to rotation is stratification or buoyancy. For example, Rayleigh-Bénard convection, where buoyancy is the primary driver, and Taylor-Couette flow, where rotation is the dominant factor, have the similar ultimate scaling of heat or angular momentum transport and Reynolds number \cite{Huisman2012,Zhu_Stevens_Shishkina_Verzicco_Lohse_2019,Jiang2022}. Their turbulent statistics and flow structures are also similar when their transport coefficients are chosen to be consistent\cite{Brauckmann2017}. \citet{Zhang_Sun_2024} showed the similarity between rotation and stratification on wall turbulence, where the Coriolis and buoyancy forces are ‘twin forces’ whose effects can be unified.
The analogy between rotation and stratification inspires us to construct an isotropic rotating system by employing appropriate stratification in conjunction with rotation.



The structure of the rest of the papers is as follows. In section \ref{Gov}, we introduce an idealized isotropic rotating system which perfectly satisfies Zeman's and Zhou's assumptions. 
By asymptotic expansion, we obtain the equation for the evolution of wave amplitude in section \ref{Asym}. 
In section \ref{Spe}, we predict the existence of $K^{-2}$ spectrum from the amplitude equation by dimensional analysis under the assumption of strong turbulence. 
In section \ref{Num}, numerical simulations of the amplitude equation and original equations both show $K^{-2}$ spectrum. 
The differences between the current isotropic system and the classic anisotropic system are also studied.
We summarize and discuss our results in section \ref{Dis}. We show the consistent spectra of high-order terms in Appendix \ref{Order1} and discuss the $K^{-2}$ spectrum by using the wave turbulence theory in Appendix \ref{WT}.

\section{\label{Gov}Governing equations}

We consider an idealized isotropic rotating stratified system proposed by \citet{xie2019two}, which stems from the inviscid two-dimensional ($\partial_y=0$) rotating stratified Boussinesq equations \cite{2019VallisBook}
\begin{subequations}
	\begin{align}
		u_t + u u_x + w u_z - f v & = -p_x,\\
		v_t + u v_x + w v_z + f u & = 0,\\
		w_t + u w_x + w w_z - b   & = -p_z,\\
		b_t + u b_x + w b_z + N^2 w & = 0,\\
		u_x + w_z & = 0,
	\end{align}\label{eq:oorig}
\end{subequations}
where $u$, $v$ and $w$ are the velocities in the $(x,y,z)$ directions, $f=2\Omega$ is the Coriolis frequency, $b$ is the buoyancy and $N$ is the buoyancy frequency. Defining $\theta=b/N$ and introducing a streamfunction $\psi$ such that $(u,w)=\nabla^\perp\psi$ with $\nabla^\perp=(-\pa{z},\pa{x})$, equations (\ref{eq:oorig}) in a special \textit{isotropic} case with $f=N$ can be simplified to
\begin{subequations}
	\begin{align}
		\nabla^2\psi_t+\Ja{\psi}{\nabla^2 \psi}-f \theta_x+ f v_z=0,\\
		v_t+\Ja{\psi}{v} -f\psi_z=0,\\
		\theta_t+\Ja{\psi}{\theta} +f\psi_x=0,
	\end{align}\label{eq:isof}
\end{subequations}
where $\Ja{a}{b}=a_x b_z-a_z b_x$. 

Introducing the rescaling 
\begin{subequations}
	\begin{align}
		t\sim 1/f, x\sim L,\, z\sim L,\, v \sim U,\, \theta \sim U ,\, \psi\sim U L,
	\end{align}
\end{subequations}
(\ref{eq:isof}) is nondimensionalized to
\begin{subequations}
	\begin{align}
		\nabla^2\psi_t+Ro\Ja{\psi}{\nabla^2 \psi}- \theta_x+  v_z=0,\label{eq:isopsi}\\
		v_t+Ro\Ja{\psi}{v} -\psi_z=0,\label{eq:iso_v}\\
		\theta_t+Ro\Ja{\psi}{\theta} +\psi_x=0,\label{eq:iso_b}
	\end{align}\label{eq:iso}
\end{subequations}
where $Ro=U/fL$ is the Rossby number, capturing the relative strength between the advection and Coriolis effects. 
Notably, equations (\ref{eq:iso_v}) and (\ref{eq:iso_b}) are similar, which shows the similarity between rotation and stratification, and $v$ enters the equation (\ref{eq:isopsi}) similarly to $\theta$, therefore, we refer to $v$ and $\theta$ as buoyancy scalars and define $\mathcal{E}_p =\int\br{\abs{v}^2+\abs{\theta}^2}/2\,\,\mdd x\mdd z$ as the potential energy. 
Then the kinetic energy is defined as $\mathcal{E}_k=\int\abs{\nabla\psi}^2/2\,\,\mdd x\mdd z$. 
System (\ref{eq:iso}) preserves a total energy 
\begin{equation}
	\mathcal{E}_{\text{tot}}=\mathcal{E}_k+\mathcal{E}_p=\int\frac{1}{2}\br{\abs{\nabla\psi}^2+\abs{v}^2+\abs{\theta}^2}\mdd x\mdd z.
\end{equation}

To justify the isotropy of the system (\ref{eq:iso}), we consider a rotation transformation
\begin{subequations}
	\begin{align}
		x'=x\cos{\alpha}-z \sin{\alpha},\\
		z'=x\sin{\alpha}+z\cos{\alpha}.
	\end{align}\label{eq:rot tran}
\end{subequations}
Letting 
\begin{equation}
	\psi'\br{x',z'}=\psi\br{x,z}, \label{eq:rot def}
\end{equation}
and defining new scalars as
\begin{equation}
	v'=v\cos{\alpha}-\theta\sin{\alpha}\quad \textrm{and}\quad \theta'=v\sin{\alpha}+\theta\cos{\alpha}, \label{eq:rot sca}
\end{equation}
the system (\ref{eq:iso}) is invariant.

Rotating stratified fluid consists of vortical and wave motions, and we consider their linear decomposition based on the dispersion relation. Considering a small perturbation to a zero state, we obtain linear equations
\begin{subequations}
	\begin{align}
		\nabla^2\psi'_t- \theta'_x+  v'_z=0,\\
		v'_t -\psi'_z=0,\\
		\theta'_t +\psi'_x=0,
	\end{align}
\end{subequations}
where the characters with a prime denote perturbation variables. 
Considering a normal mode, i.e., $\psi'=\tilde{\psi'}e^{-\mi\omega t+\mi \br{k x+m z}}$, where $k$ and $m$ are the horizontal and vertical wavenumbers, respectively, we obtain the dispersion relation
\begin{equation}
	\omega_*=0 \quad \mathrm{and} \quad \omega_\pm=\pm 1, 
\end{equation}
and the corresponding eigenvectors
\begin{align}
	\begin{bmatrix}
		\psi_*\\
		v_*\\
		\theta_*
	\end{bmatrix}
	=
	\begin{bmatrix}
		0\\
		k\\
		m
	\end{bmatrix} \quad \mathrm{and} \quad
	\begin{bmatrix}
		\psi_\pm\\
		v_\pm\\
		\theta_\pm
	\end{bmatrix}
	=
	\begin{bmatrix}
		1\\
		\pm m\\
		\mp k
	\end{bmatrix}.\label{eq:chara}
\end{align}
Here, $_*$ and $_\pm$ denote the vortical and wave components, respectively. 
It is interesting to note that $\psi$ is only related to the wave part, which we use in section \ref{Oe} to avoid exciting vortical models in numerical simulations. 

This isotropic system (\ref{eq:iso}) with wavenumber-independent dispersion relation perfectly satisfies Zeman's and Zhou's assumptions and choices of a single time scale for the rotation effect.
In addition, the nondimentionalized frequency takes two definite values $\omega=\pm 1,$ corresponding to Coriolis frequency, therefore triad interaction of linear wave is not permitted.
In comparison, the current isotropic scenario is much simpler than the anisotropic case with $f\neq N$, where the linear dispersion
relation for inertia–gravity waves is \cite{2019VallisBook} 
\begin{equation}
	\omega=\pm \sqrt{\frac{N^2 k^2+f^2 m^2}{k^2+m^2}},
\end{equation}
which is anisotropic and could permit a triad wave interaction \cite{Nazarenko2011wave}. 

\section{\label{Asym}Amplitude equation}
To elucidate the nonlinear interaction of weak linear waves and to obtain the corresponding energy spectra of our system (\ref{eq:iso}) under the rapid rotation condition, we derive the slow evolution equation of the wave amplitude by asymptotic expansion under $Ro\ll 1$.

\subsection{Reformulating the system (\ref{eq:iso})}

We first introduce a Helmholtz decomposition to the buoyancy scalars
\begin{equation}
	(v,\theta)=\nabla \Phi+\nabla^{\perp}\Psi.
\end{equation} 
Therefore, the polarization relation (\ref{eq:chara}) leads to the constraints on the linear vortical flow ($\omega_*=0$) 
\begin{equation}
	\psi=0, \quad \nabla^2 \Psi=0,
\end{equation}
and on the linear wave ($\omega_\pm=\pm 1$)
\begin{equation}
	\nabla^2 \Phi=0.
\end{equation}
Thus, considering a suitable boundary condition, e.g., a doubly periodic boundary condition, at the linear level, $\psi$ and $\Psi$ capture the wave component, and $\Phi$ describes the vortical flow. 

Rewriting equations (\ref{eq:iso}) in terms of $\psi$, $\Psi$ and $\Phi$, we obtain
\begin{subequations}
\begin{align}
    \nabla^2 \psi_t-\nabla^2 \Psi +{Ro}\Ja{\psi}{\nabla^2 \psi}=0,\\
    \nabla^2 \Psi_t+\nabla^2 \psi +{Ro}\left( \Ja{\psi}{\nabla^2 \Psi}+\Ja{\psi_x}{\Psi_x}+\Ja{\psi_z}{\Psi_z}+\Ja{\psi_x}{\Phi_z}-\Ja{\psi_z}{\Phi_x}\right)=0,\\
    \nabla^2 \Phi_t+{Ro}\left(-\Ja{\psi_x}{\Psi_z}+\Ja{\psi_z}{\Psi_x}+\Ja{\psi}{\nabla^2 \Phi}+\Ja{\psi_x}{\Phi_x}+\Ja{\psi_z}{\Phi_z}\right)=0,
\end{align}\label{eq:iso2}
\end{subequations}
and the total energy becomes  
\begin{equation}
    \mathcal{E}_{\text{tot}}=\int\frac{1}{2}\br{\abs{\nabla\psi}^2+\abs{\nabla\Phi}^2+\abs{\nabla\Psi}^2}\mdd x\mdd z,\label{eq:energy2}
\end{equation}
which consists of the kinetic energy of waves $\int\abs{\nabla\psi}^2/2\,\,\mdd x\mdd z$ , the potential energy of waves $\int\abs{\nabla\Psi}^2/2\,\,\mdd x\mdd z$ and the energy of vortices $\int\abs{\nabla\Phi}^2/2\,\,\mdd x\mdd z$.

\subsection{Asymptotic expansion}

We consider rapidly rotating scenario with $ Ro \ll 1$, which leads to the dominant linear effect and the nonlinear effects are of high order. 
So we introduce asymptotic expansions 
\begin{subequations}
	\begin{align}
		\psi=\psi_0+Ro \psi_1+ Ro^2 \psi_{2}+\dots,\\
		\Psi=\Psi_0+Ro \Psi_1+Ro^2 \Psi_{2}+\dots,\\
		\Phi=\Phi_0+Ro \Phi_1+Ro^2 \Phi_{2}+\dots.
	\end{align}\label{eq:expand}
\end{subequations}
To explore the rotating dominant scenario, we focus on the wave-dominant case, i.e., $\Phi_0=0$. Substituting the expansions (\ref{eq:expand}) into the governing equations (\ref{eq:iso2}), we obtain equations of various orders of $Ro$.

At $O(1)$, we obtain
\begin{subequations}
	\begin{align}
		\partial_t \nabla^2 \psi_0-\nabla^2 \Psi_0 =0,\\
		\partial_t \nabla^2 \Psi_0+\nabla^2 \psi_0 =0.
	\end{align}
\end{subequations}
Whereafter, one obtain the leading-order solution 
\begin{equation}
	\psi_0+\mi\Psi_0=A \me^{-\mi t},\label{eq:order0}
\end{equation}
where $A(x,z,\tau)$ is the wave amplitude, a complex function of slow time with $\tau=Ro^2 t$. 
Here, the slow time set to be $O(Ro^2)$, which we explain below. 

The $O(Ro)$ equations are 
\begin{subequations}
	\begin{align}
		\nabla^2 \psi_{1t}-\nabla^2 \Psi_1 +\left(\frac{1}{4}\Ja{A}{\nabla^2 A} \me^{-2\mi t}+\frac{1}{4}\Ja{A}{\nabla^2 A^{*}}+c.c.\right)&=0,\\
		\nabla^2 \Psi_{1t}+\nabla^2 \psi_1+\qquad\qquad\qquad\qquad\qquad\qquad\qquad\qquad\qquad\qquad & \nonumber\\
		\bigg(-\frac{\mi}{4}\Ja{A}{\nabla^2 A}\me^{-2\mi t}+\frac{\mi}{4}\Ja{A}{\nabla^2 A^*}+\frac{\mi}{4}\Ja{A_x}{A^*_x}
		+\frac{\mi}{4}\Ja{A_z}{A^*_z}+c.c.\bigg)&=0,\\
		\nabla^2 \Phi_{1t}+\left(\frac{\mi}{2}\Ja{A_x}{A_z}\me^{-2\mi t}+c.c.\right)&=0,
	\end{align}\label{eq:asym oe}
\end{subequations}
where $c.c.$ denotes the complex conjugate. 
(\ref{eq:asym oe}) are solvable with solutions
\begin{subequations}
	\begin{align}
		\nabla^2 \psi_{1}=-\frac{\mi}{4}\Ja{A}{\nabla^2 A}\me^{-2\mi t}-\frac{\mi}{4}\nabla\cdot\Ja{A}{\nabla A^*}+c.c.,\label{eq:order1psi}\\
		\nabla^2 \Psi_{1}=-\frac{1}{4}\Ja{A}{\nabla^2 A}\me^{-2\mi t}+\frac{1}{4}\Ja{A}{\nabla^2 A^*}+c.c.,\\
		\nabla^2 \Phi_{1}=\frac{1}{4}\Ja{A_x}{A_z}\me^{-2\mi t}+c.c..
	\end{align}\label{eq:order1}
\end{subequations}
In this order, if we introduce a slow time $\tau'=Ro t$, the first two equations in (\ref{eq:asym oe}) would introduce a solvability condition of $A_{\tau'}=0$, indicating a slower time scale. 

At $O(Ro^2)$, we obtain
\begin{subequations}
	\begin{align}
		\partial_t \nabla^2 \psi_{2}-\nabla^2 \Psi_2 +\partial_{\tau}\nabla^2\psi_0+\underbrace{\Ja{\psi}{\nabla^2 \psi}|_{0,1}}_{\mathcal{J}_1(A) \me^{-\mi t}+\mathcal{J}_1(A)^* \me^{\mi t}}&=0,\label{eq:psi2}\\
		\partial_t \nabla^2 \Psi_{2}+\nabla^2 \psi_2 +\partial_{\tau}\nabla^2\Psi_0+\qquad\qquad\qquad\quad\notag\\\underbrace{\left( \Ja{\psi}{\nabla^2 \Psi}+\Ja{\psi_{x}}{\Psi_{x}}+\Ja{\psi_{z}}{\Psi_{z}}+\Ja{\psi_{x}}{\Phi_{z}}-\Ja{\psi_{z}}{\Phi_{x}}\right)|_{0,1}}_{\mathcal{J}_2(A) \me^{-\mi t}+\mathcal{J}_2(A)^* \me^{\mi t}}&=0,\label{eq:PPsi2}
	\end{align}\label{eq:order2eq}
\end{subequations}
where the lower index $|_{0,1}$ denotes the interaction of $_0$ and $_1$ components in the quadratic terms. 
Note that we the equation for $\Phi_2$ is solvable, its equation is omitted in the above equations (\ref{eq:order2eq}).

Substituting the leading- and first-order solutions, (\ref{eq:order0}) and (\ref{eq:order1}), into equations (\ref{eq:order2eq}), and using $\mathcal{J}_1(A) \me^{-\mi t}+\mathcal{J}_1(A)^* \me^{\mi t}$ and $\mathcal{J}_2(A) \me^{-\mi t}+\mathcal{J}_2(A)^* \me^{\mi t}$ to denote cross terms in (\ref{eq:psi2}) and (\ref{eq:PPsi2}), respectively, we obtain
\begin{subequations}
\begin{align}
    \partial_t \nabla^2 \psi_{2}-\nabla^2 \Psi_2 +\br{\me^{-\mi t}\left(\frac{1}{2}\nabla^2 A_{\tau}+\mathcal{J}_1(A)\right)+c.c.}=0,\label{eq:psi20}\\
    \partial_t \nabla^2 \Psi_{2}+\nabla^2 \psi_2 +\br{\me^{-\mi t}\left(\frac{1}{2\mi}\nabla^2 A_{\tau}+\mathcal{J}_2(A)\right)+c.c.}=0.\label{eq:PPsi20}
\end{align}
\end{subequations}
Multiplying equation (\ref{eq:PPsi20}) by $\mi$ and adding it to equation (\ref{eq:psi20}), we obtain
\begin{align}
    \partial_t \nabla^2 \left(\psi_2+\mi\Psi_2\right)+\mi \nabla^2 \left(\psi_2+\mi\Psi_2\right)+\br{\me^{-\mi t}\left(\nabla^2 A_{\tau}+\mathcal{J}_1(A)+\mi\mathcal{J}_2(A)\right)+c.c.}=0.
    \label{eq:epsilon2}
\end{align}

The solvability condition for $\psi_2+\mi \Psi_2$ in equation (\ref{eq:epsilon2}) results in the equation for the slow evolution of wave amplitude $A(x,z,\tau)$:
\begin{equation}
	\nabla^2 A_{\tau}+\frac{\mi}{4}J_{\text{NL}}=0,\label{eq:wave amplitude} 
\end{equation}
with
\begin{equation}
	\begin{aligned}
		J_{\text{NL}}=&-\Ja{A}{\Jacd{\nabla A}{\nabla A^*}}\\
		&+\Jacd{\nabla A}{\nabla\Delta^{-1}\nabla\cdot\Ja{A}{\nabla A^*}}
		-\Jacd{\nabla A^*}{\nabla\Delta^{-1}\Ja{A}{\nabla^2 A}}\\
		&+\frac{1}{2}\Ja{A_x^*}{\partial_z\left[\Delta^{-1}\Ja{A_x}{A_z}\right] }
		-\frac{1}{2}\Ja{A_z^*}{\partial_x\left[\Delta^{-1}\Ja{A_x}{A_z}\right] },
	\end{aligned}
\end{equation}
where $\Jacd{\nabla a}{\nabla b}=\Ja{a_x}{b_x}+\Ja{a_z}{b_z}$ and $\Delta^{-1}$ is the inverse of Laplacian operator. 
The nonlinear terms show that wave quartic interaction controls the slow dynamics, which is also a corollary of the dispersion relation $\omega=\pm 1$.
It is special that all nonlinear terms in the above amplitude equation contain the Jacobian operator, therefore, a one-dimensional correspondence of the current system is not possible because a $x$ (or $z$)-independent solution has no nonlinear effect.
This form distinguishes it from other wave systems such as the MMT system\cite{Majda1997AOM}.

Multiplying equation (\ref{eq:wave amplitude}) by $A^*$, adding the conjugate of the resultant equation and integrating over space with doubly periodic boundary conditions, we obtain the energy conservation
\begin{align}
	\partial_{\tau}\int \nabla A\cdot\nabla A^*\mdd x\mdd z=0.
\end{align}
Considering that at the the leading order $\Phi_0=0$, the vortex energy $\int\abs{\nabla\Phi}^2/2\,\,\mdd x\mdd z$ is of high order. 
Thus, the total energy (\ref{eq:energy2}) is approximated as
\begin{equation}
	\begin{aligned}
	\mathcal{E}_{\text{tot}}&=\int\frac{1}{2}\br{\abs{\nabla\psi}^2+\abs{\nabla\Psi}^2}\mdd x\mdd z \\ &\approx\int\frac{1}{2}\br{\abs{\nabla\psi_0}^2+\abs{\nabla\Psi_0}^2}\mdd x\mdd z=\int\frac{1}{2}\abs{\nabla A}^2\mdd x\mdd z \equiv \mathcal{E}_{A}.
	\end{aligned}
\end{equation}


\subsection{\label{Spe} Energy spectrum}

We can obtain the wave energy spectrum from the nonlinear wave amplitude equation (\ref{eq:wave amplitude}). 
Since the quartic interaction with constant frequency does not rule out any interacting modes, which happens in dispersive wave turbulence, we follow the strong turbulence argument \cite[cf.][]{kolmogorov1941dissipation}.
Denoting $L$ as a length scale, the nonlinear terms of wave amplitude equation (\ref{eq:wave amplitude}) scale as, e.g., 
\begin{equation}
	J_{\text{NL}}\sim \Ja{A}{\Jacd{\nabla A}{\nabla A^*}}\sim A^3/L^6,
\end{equation}
and the other nonlinear terms have the same dimension. 
Thus, the strong turbulence implies that the energy flux scales as $AJ_{\text{NL}} \sim A^4/L^6$.
Considering Kolmogorov's constant-flux picture, we have $A^4/L^6=const$, i.e., $A\sim K^{-3/2}$ with $K=\sqrt{k^2+m^2}$.
Since
\begin{equation}
	\int E(K)\mdd K=\frac{\int \abs{\nabla A}^2 \mdd \bs{x}}{\int \mdd \bs{x}},
\end{equation}
the energy spectrum follows
\begin{equation}
	E(K)\sim K^{-2},
\end{equation}
which is consistent with the result of \citet{Zeman1994} and \citet{Zhou1995}.
Additionally, we notice that, as mentioned in our hypothesis, linear waves dominate this system, therefore we may use wave turbulence theory to study this problem, which we discuss in appendix \ref{WT}.

\section{\label{Num}Numerical simulations}

We performed forcing-dissipative numerical simulations of both the amplitude equation (\ref{eq:wave amplitude}) and original equations (\ref{eq:iso}) with zero initial state. The simulations are performed in a box of size $2\pi\times 2\pi$ with a resolution $512\times512$. 
We use a Fourier pseudospectral method with $1/2$ dealiasing for amplitude equation and $2/3$ dealiasing for original equations in space, and a fourth-order explicit Runge-Kutta scheme in time. 

\citet{xie2019two} showed a forward energy cascade in the original system (\ref{eq:iso}).
Therefore, we can exert force from the very first wavenumber. 
The white-noise forcing term, which we will specify in equations (\ref{eq:simAeq}) and (\ref{eq:simoeq}) below, is prescribed in the Fourier space by
\begin{equation}
	F = \Re\{ \ifourier{\hat{F}} \} \quad \mathrm{with} \quad \hat{F}=M_F H(K_f-K)\mathcal{A} \me^{\mi 2\pi \mathcal{B}},\label{eq:forcing}
\end{equation}
where $M_F$ is the magnitude of forcing, $H$ is the Heaviside funtion, $\mathcal{A}$ is a Gaussian random variable and $\mathcal{B}$ follows a uniform distribution. 
To ensure enough wave nonlinear interaction, the forcing acts in a wide range $[1,\,K_f]$ in the wavenumber space. 


Following \cite{smith_2002}, we choose an exponential cutoff filter for dissipation,
\begin{equation}
	H_{ec}\br{K}=\left\{
	\begin{aligned}
		&\text{exp}\br{-a\br{K-K_{\text{cut}}}^s},\quad &K> K_{\text{cut}}\\
		&1,&K\le K_{\text{cut}}
	\end{aligned},
	\right.
\end{equation}
where 
\begin{equation}
	a=\frac{\ln \br{1+4\pi/N}}{\br{K_{\text{max}}-K_{\text{cut}}}^s}.
\end{equation}
Here $N$ is the resolution, $K_{\text{max}}$ is the maximum wavenumber, $K_{\text{cut}}$ is the cutoff wavenumber, and $s$, which we choose to be $2$, is a parameter describing the order of a corresponding hyperviscosity filter.
This filter acts on the equation via $\hat{\psi}_n=\hat{\psi}_n^{'} H_{ec}\br{K}$, where $\hat{\psi}_n^{'}$ is the Fourier transform of $\psi$ calculated by the non-dissipative equation at the $n$th time step. 
This process can be replaced by a pseudo dissipative term \cite[cf.][]{smith_2002} $D\psi$ added to the right-hand side of the equation.
Since there is no dissipation at wavenumbers smaller than the cutoff wavenumber, $K_{\text{cut}}$, we can obtain a range in the wavenumber space with a constant energy flux.

\subsection{\label{Amp}The amplitude equation}

In this section, we show the numerical results of the amplitude equation (\ref{eq:wave amplitude}). 
With forcing and dissipation, the equation becomes 
\begin{equation}
	\nabla^2 A_{\tau}=-\frac{\mi}{4}J_{\text{NL}}+ \ifourier{\hat{F}} + DA,\label{eq:simAeq}
\end{equation}
where $F$ is the forcing with $K_f=4$, $M_F=400$ (cf. (\ref{eq:forcing})), and energy injection rate $\epsilon=2.3\times 10^{-6}$, and $D$ is the equivalent form of exponential cutoff dissipation with $K_{\text{cut}}=160$.


The evolution of total energy $\mathcal{E}_A=\int \nabla A\cdot\nabla A^*/2\,\mdd x\mdd z$ reaches the statistically steady phase around $\tau=7000$ and the statistical quantities are calculated among ($\tau\in[7000,\,8000]$). 

\begin{figure}
	\centering
	\includegraphics[width=0.47\textwidth]{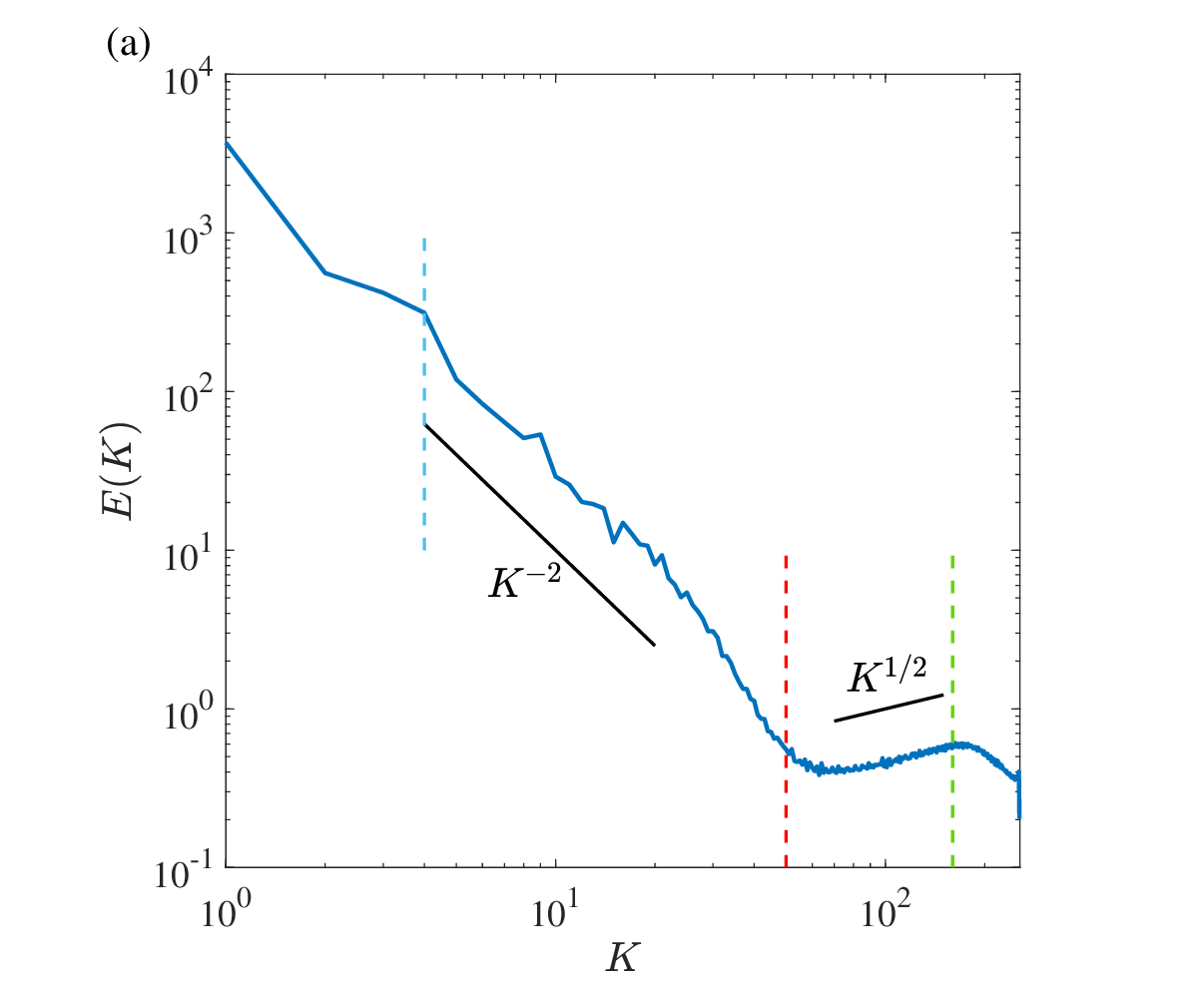}
	\includegraphics[width=0.47\textwidth]{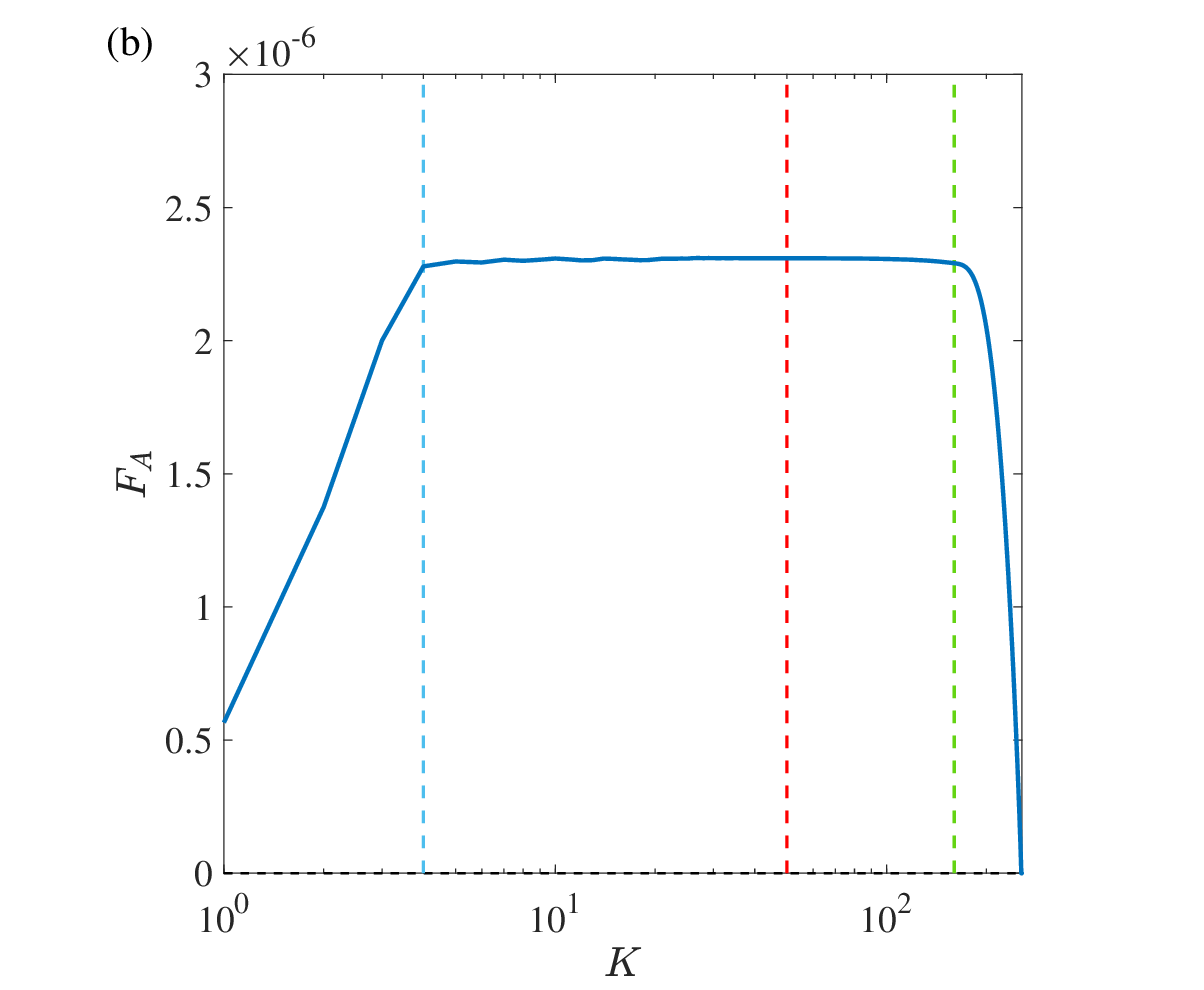}
	\caption{Energy spectrum $E_A(K)$ (a) and energy flux (b) (positive means downscale) averaged over $\tau\in[7100,7900]$ of the simulation with $K_f=4$ and $\epsilon=2.3\times 10^{-6}$. The blue, red and green dashed lines mark $K_{f}=4$, $K=50$ and $K_{\text{cut}}=160$, respectively. The two solid black lines are references for $K^{-2}$ and $K^{1/2}$. }
	\label{fig:aespeflux}
\end{figure}
%
Fig. \ref{fig:aespeflux}(a) shows the energy spectrum $E_A(K)$, where the notation of $E$ denote the energy spectrum, and for a given field $\chi$, $\int E_{\chi}(K) \mdd K= \int \nabla \chi\cdot\nabla \chi^*/2\,\mdd x \mdd z $. 
The spectrum approximately proportional to $K^{-2}$ in the range $K\in[4,50]$ and $K^{1/2}$ in the range $K\in [60,160]$. 

We calculate the total energy flux by
\begin{equation}
	F_A(K)= \int_{0}^{K}\int_{0}^{2\pi} \frac{1}{2}\br{-\frac{\mi}{4}\hat{A}^*\widehat{J_{\text{NL}}} + c.c. } \kappa \mdd \kappa \mdd \alpha,
\end{equation}
where $\hat{\cdot}$ denotes the Fourier transform, 
and we show total energy flux in Fig. \ref{fig:aespeflux}(b). We observe an inertial range of constant energy flux between the forcing wavenumber $K_f=4$ and the dissipation wavenumber $K_{\text{cut}}=160$.
The existence of the inertial range and the $K^{-2}$ energy spectrum aligns with our deduction in section \ref{Spe}. 
Concerning the 'bottleneck' range $K\in [60,160]$ in Fig. \ref{fig:aespeflux}(a), the $K^{1/2}$ scaling could be a combination of
the Kolmogorov-Zakharov spectrum and the Rayleigh-Jeans spectrum \cite{Berhanu2022}. Using the spectrum of $E_A$ and equations (\ref{eq:order0})(\ref{eq:order1}), we can deduce by dimension analysis the spectra of $O(Ro)$ terms, which we discuss in Appendix \ref{Order1ae}.

\subsection{\label{Oe}The original equations}
In this section, we perform numerical simulations of the original equations (\ref{eq:iso}). 
Here, we emphasize that the forcing is only exerted on the $\psi$ equation (\ref{eq:isopsi}) to avoid the direct excitation of the vortical part (cf. (\ref{eq:chara})). 
With forcing and dissipation, the equations become
\begin{subequations}
	\begin{align}
		\nabla^2\psi_t+\Ja{\psi}{\nabla^2 \psi}-\theta_x+v_z &= F + D\psi,\\
		v_t+\Ja{\psi}{v} -\psi_z &= Dv,\\
		\theta_t+\Ja{\psi}{\theta} +\psi_x &= D\theta,
	\end{align}\label{eq:simoeq}
\end{subequations}
where $F$ is the forcing with $K_f=4$, $M_F=3$ (cf. (\ref{eq:forcing})), and $D$ is the equivalent form of exponential cutoff dissipation with $K_{\text{cut}}=150$. 

Note that we omit $Ro$ in the nonlinear terms.
By dimensional analysis, the energy injection rate $\epsilon\sim U^3/L$ and Zeman scale, where the characteristic scale of inertial wave and eddy turn-over time are equal, $L_Z\sim\epsilon^{1/2}f^{-3/2}$, then $Ro=U/fL\sim\epsilon^{1/3}L^{-2/3}f^{-1}=(L_Z/L)^{2/3}=(K/K_Z)^{2/3}$, where $K_Z$ is Zeman wavenumber. Thus $Ro\ll 1$ 
is equivalent to $K\ll K_Z$ here.
Therefore, in the current simulation, we design $K_Z$ to be much bigger than the wavenumbers in the inertial range to guarantee the rapidly rotating turbulence by choosing proper values of $M_F$.

In the current simulation, the total energy injection rate is $\epsilon=1.1\times 10^{-8}$ and the corresponding Zeman wavenumber is $K_Z= \epsilon^{-1/2} f^{3/2}\approx 9.4\times 10^3$ (here $f=1$ due to our rescaling), which is much larger than $K_{\text{cut}}$ and ensures the dominance of linear waves and a rapidly rotating turbulence. 
\begin{figure}
	\centering
	\includegraphics[width=0.5\textwidth]{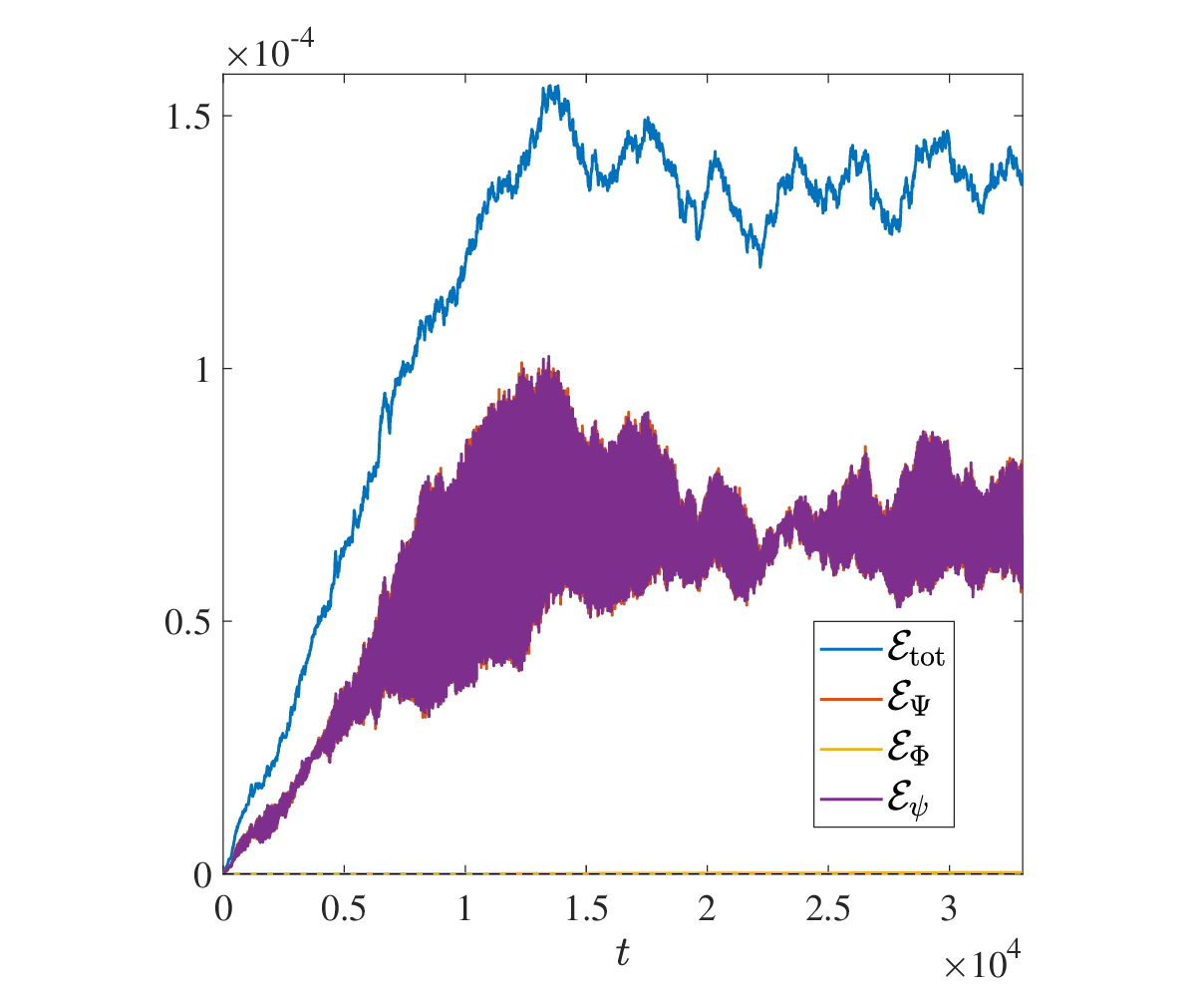}
	\caption{Evolution of wave kinetic energy $\mathcal{E}_{\psi}$, wave potential energy $\mathcal{E}_{\Phi}$, vortical energy $\mathcal{E}_{\psi}$, and total energy $\mathcal{E}_{\text{tot}}=\mathcal{E}_{\psi}+\mathcal{E}_{\Psi}+\mathcal{E}_{\Phi}$ of the simulation with $K_f=4$ and $\epsilon=1.1\times 10^{-8}$.}
	\label{fig:oeevo}
\end{figure}
\begin{figure}
	\centering
	\includegraphics[width=0.47\textwidth]{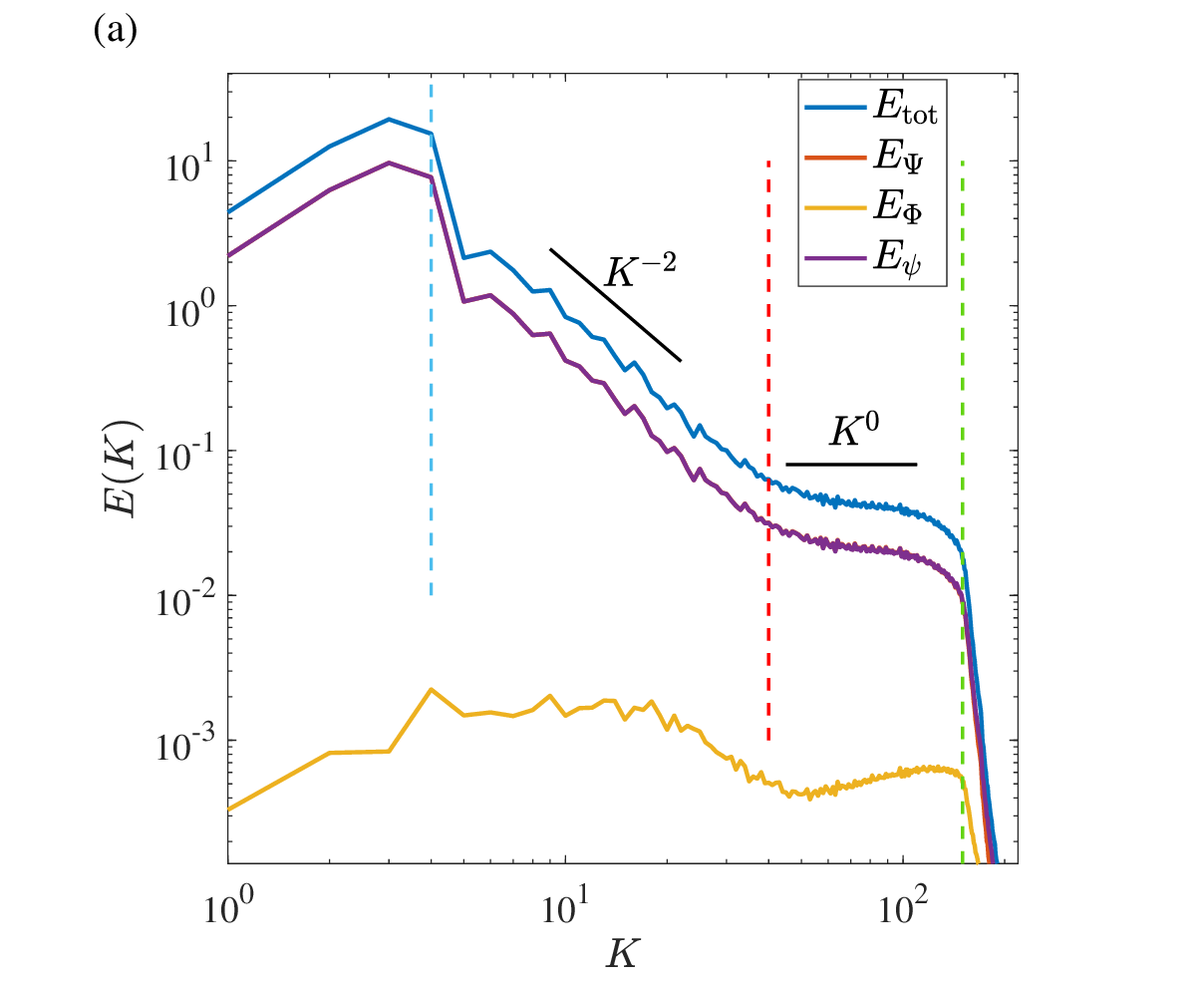}
	\includegraphics[width=0.47\textwidth]{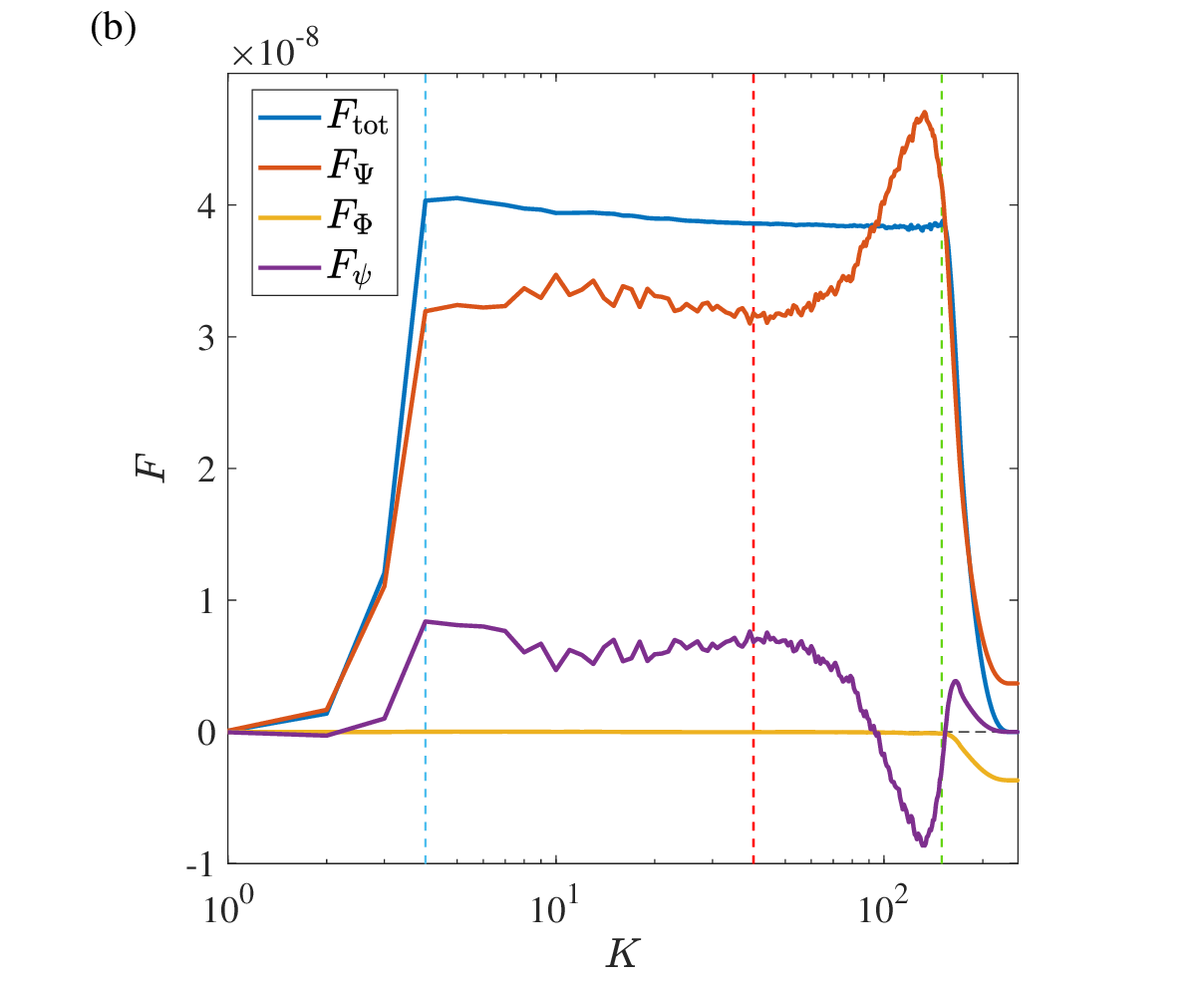}
	\caption{Energy spectrum (a) and energy flux (b) of the simulation with $K_f=4$ and $\epsilon=1.1\times 10^{-8}$. The two solid black lines are $K^{-2}$ and $K^{0}$ as references. The blue, red and green dashed lines mark $K_{f}=4$, $K=40$ and $K_{\text{cut}}=150$, respectively. $F_{\text{tot}}$ is energy flux of total energy, and $F_{\Psi}$, $F_{\Phi}$, $F_{\psi}$ are the corresponding energy flux of $E_{\Psi}$, $E_{\Phi}$ and $E_{\psi}$, respectively.}
	\label{fig:oespeflux}
\end{figure}

We plot in Fig. \ref{fig:oeevo} the evolution of wave kinetic energy $\mathcal{E}_{\psi}=\int \abs{\nabla \psi}^2/2\,\,\mdd x\mdd z$, wave potential energy $\mathcal{E}_{\Psi}=\int \abs{\nabla \Psi}^2/2\,\,\mdd x\mdd z$, and the vortical energy $\mathcal{E}_{\Phi}=\int \abs{\nabla \Phi}^2/2\,\,\mdd x\mdd z$, which is of high order in the current setup.
As expected, the vortical energy is far less than that of waves, justifying that our simulation is a good realization of a wave-dominant system. 
Potential and kinetic energy fluctuate and compensate for each other.

We show energy spectra and energy fluxes in Fig. \ref{fig:oespeflux}, where the fluxes are defined by
\begin{subequations}
	\begin{align}
		&F_{\Psi}(K)= \int_{0}^{K}\int_{0}^{2\pi} \frac{1}{2}\br{\hat{\Psi}^*\hat{N_{\Psi}} + c.c. } \kappa \mdd \kappa \mdd \alpha,\quad N_{\Psi}=\partial_x\Ja{\psi}{\theta}-\partial_z\Ja{\psi}{v},\\
		&F_{\Phi}(K)= \int_{0}^{K}\int_{0}^{2\pi} \frac{1}{2}\br{\hat{\Phi}^*\hat{N_{\Phi}} + c.c. } \kappa \mdd \kappa \mdd \alpha, \quad N_{\Phi}= \partial_z\Ja{\psi}{\theta}+\partial_x\Ja{\psi}{v},\\
		&F_{\psi}(K)= \int_{0}^{K}\int_{0}^{2\pi} \frac{1}{2}\br{\hat{\psi}^*\hat{N_{\psi}} + c.c. } \kappa \mdd \kappa \mdd \alpha,\quad N_{\psi}=-\Ja{\psi}{\nabla^2 \psi},\\
		&F_{\text{tot}}(K)=F_{\Psi}(K)+F_{\Phi}(K)+F_{\psi}(K).
	\end{align}
\end{subequations}
We observe an inertial range $K\in [4,40]$ and a corresponding  $K^{-2}$ spectrum. 
The overlapping of curves of $E_{\Psi}$ and $E_{\psi}$ shows the linear wave dominance. 
Also, we find an approximate $K^0$ scaling of the energy spectrum in the range with $K\in [40,150]$, which could be a combination of the Kolmogorov-Zakharov spectrum and the Rayleigh-Jeans spectrum.
These results are consistent with those of the amplitude equation in section \ref{Amp}.

\begin{figure}[h]
	\centering
	\includegraphics[width=0.47\textwidth]{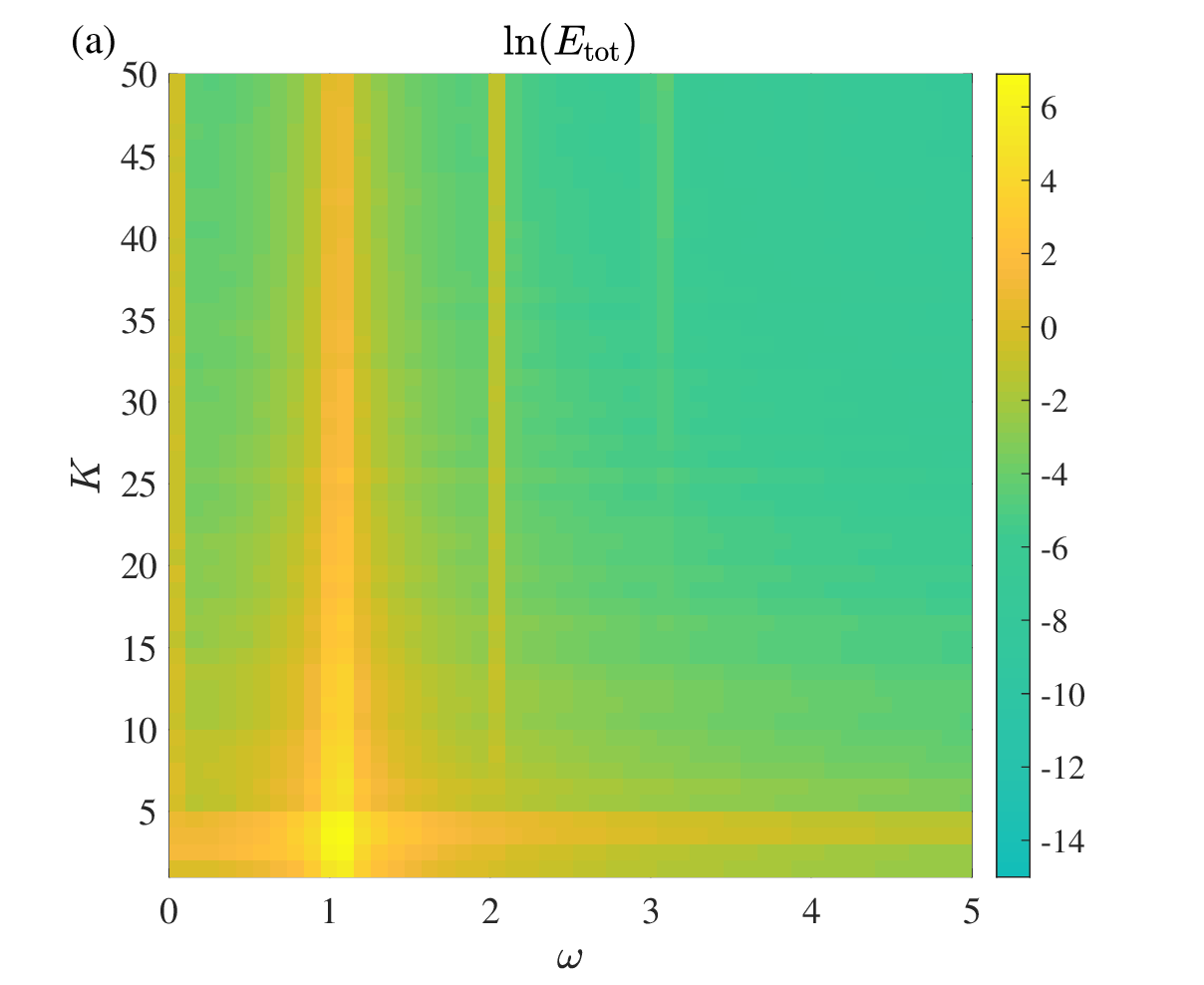}
	\includegraphics[width=0.47\textwidth]{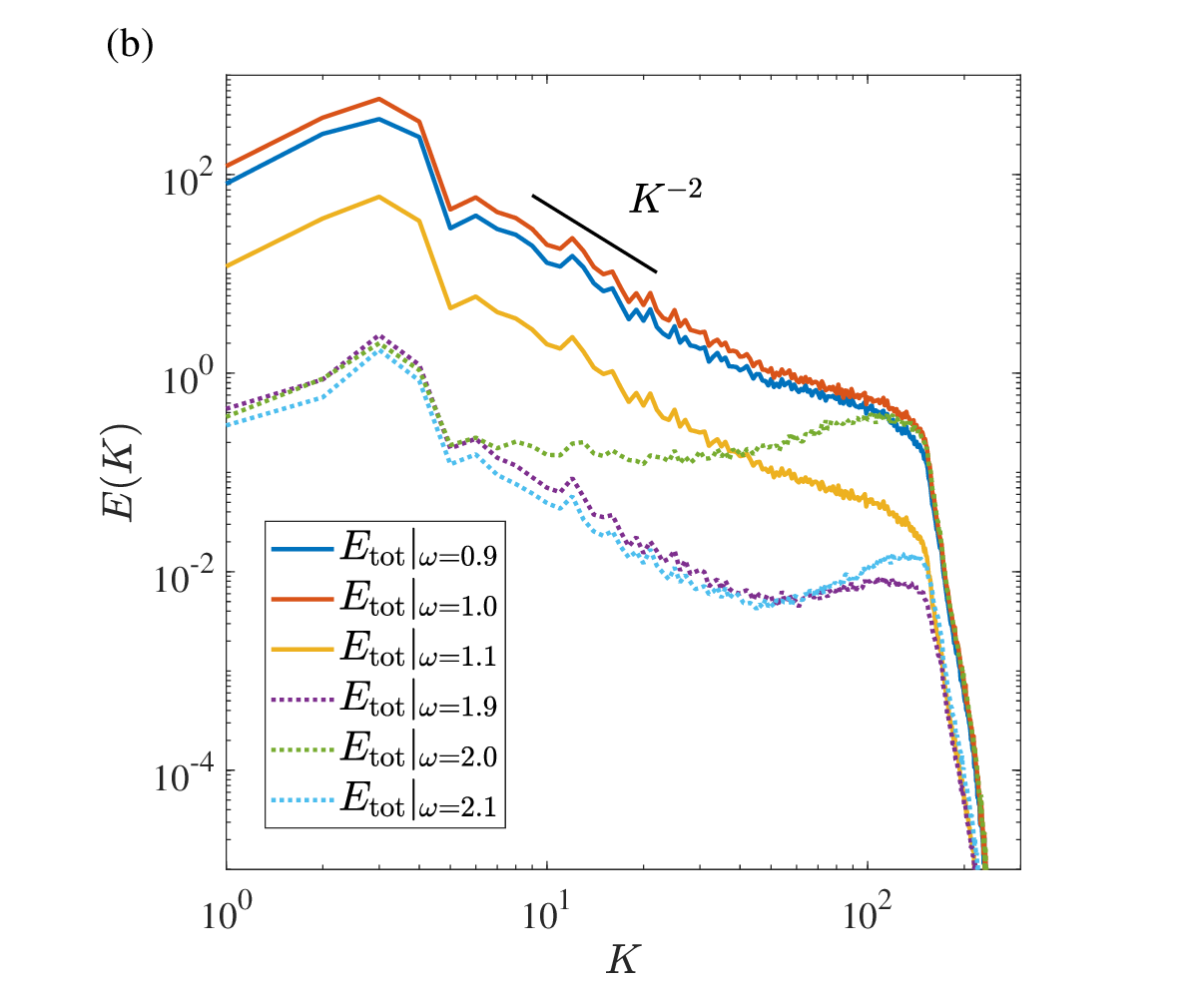}
	\caption{(a) The total spatial-temporal Fourier spectrum. (b) Spectra of total energy at different $\omega$ are plotted. The solid black line is $K^{-2}$.
	}
	\label{fig:spacetimespe}	\label{fig:seperatespe} 
\end{figure}

The spatial-temporal spectrum shown in Fig. \ref{fig:spacetimespe}(a) also justifies the weakly nonlinear character. 
The total energy $E_{\text{tot}}$ centres along the linear dispersion relation $\omega=\pm 1$. 
And the energy concentration along $\omega=2$ and $\omega=0$ accord with the wave quartic interaction. 
The energy concentration in the range $K\in[1,4]$ is a result of forcing.
The energy spectra at different frequencies $\omega$ are shown in Fig. \ref{fig:seperatespe}(b). 
Almost all the spectra scale as $K^{-2}$, while at $\omega=2.0$, $E_{\text{tot}}|_{\omega=2}\sim K^0$. Notably, $E_{\text{tot}}|_{\omega=2}$ is $O(Ro)$ and we discuss $O(Ro)$ terms in Appendix \ref{Order1oe}.

\subsection{The classic (anisotropic) 2D rotating turbulence}

As a comparison, we perform a simulation of the classic anisotropic 2D rotating turbulence, whose governing equations are
\begin{subequations}
	\begin{align}
		\nabla^2\psi_t+\Ja{\psi}{\nabla^2 \psi}+ v_z&= F^G + D\psi + \alpha \nabla^2 \psi,\\
		v_t+\Ja{\psi}{v} -\psi_z&=Dv +\alpha \nabla^2 v,
	\end{align}
	\label{eq:aniso}
\end{subequations}
where $F^G$ is a Gaussian forcing 
\begin{equation}
	F^G = \Re\{ \ifourier{\hat{F}^G} \} \quad \mathrm{with} \quad \hat{F}^G =M_F \me^{-\frac{\br{K-K_f}^2}{2D_K^2}} \mathcal{A} \me^{\mi 2\pi\mathcal{B}},
\end{equation}
$M_F=3$, $K_f=6$, $D_K=2$, and $D$ is the equivalent form of exponential cutoff dissipation with $K_{\text{cut}}=150$. 
The last terms on the r.h.s. are extra damping terms at large scales with $\alpha=0.1$. Note that the simulation of this anisotropic system has different parameters from the previous one of the isotropic system in Section \ref{Oe}. Like Two-dimensional Rayleigh-B\'{e}nard convection, this anisotropic system is inevitable to reach a statistically steady state without large-scale coherent structures, and we need a large scale dissipative term to saturate the inverse cascade when present (c.f. \citet{winchester2023twodimensional}).
So, to avoid this energy accumulation, we change the forcing term and add an extra damping term at large scales. 
This system (\ref{eq:aniso}) has a dispersion relation $\omega=\pm m/K \in [0,\,1]$, which implies a wide range of characteristic timescales. 
When $m\to 0$ with fixed $k$, $\omega\to 0$ and therefore $t_{wave}=1/\omega \gg t_{NL}=\epsilon^{-1/3}K^{-2/3}$, which implies the dominance of nonlinear effects, and hence like a $beta$-plane turbulence \cite{Vallis1993} energy tends to accumulate at large scale around $m=0$. 
Since the new forcing and extra damping are isotropic, they do not directly introduce anisotropic effects to the system (\ref{eq:aniso}), and the observed anisotropy should be the effect of the system itself.

\begin{figure}
	\centering
	\includegraphics[width=0.47\textwidth]{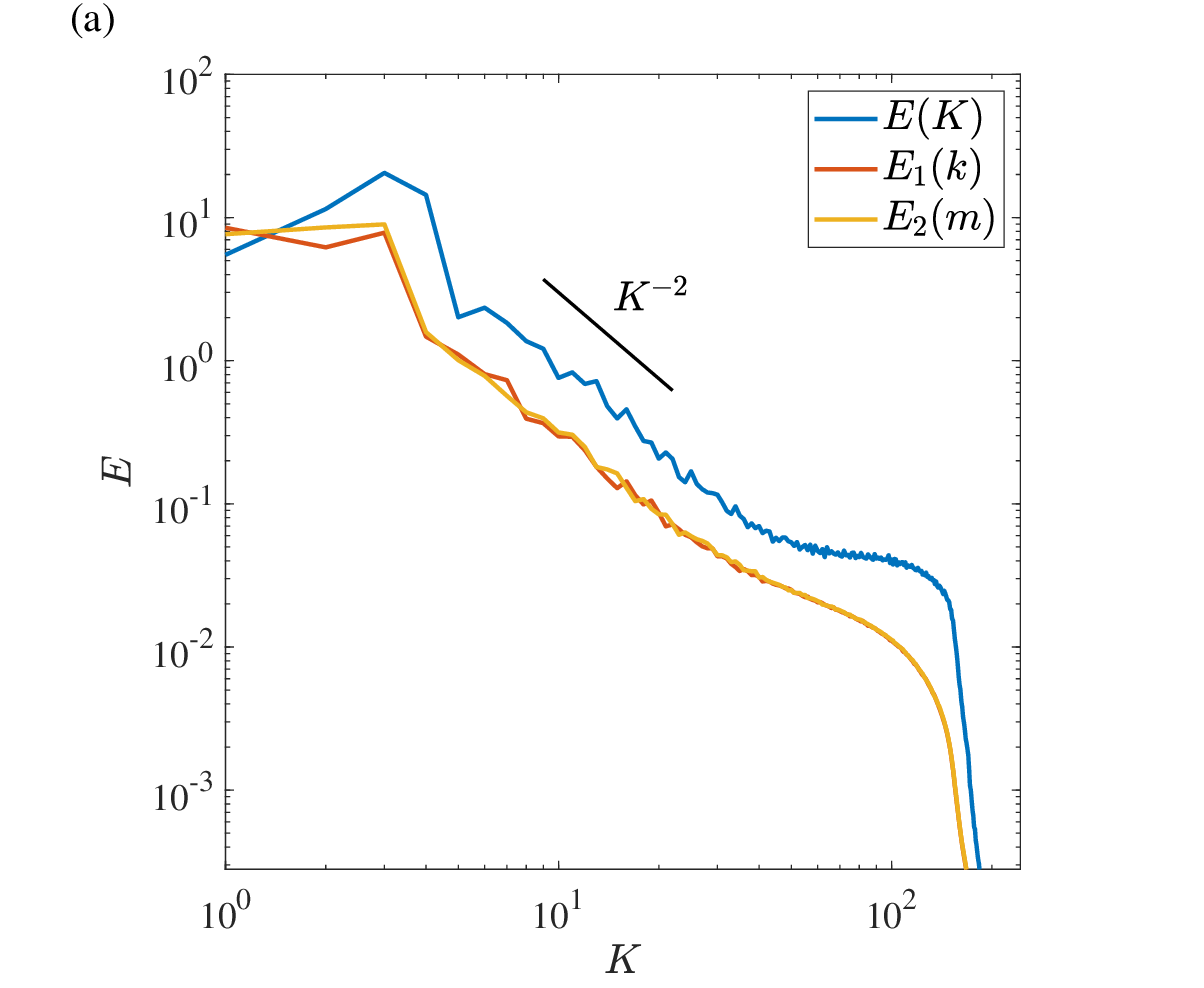}
	\includegraphics[width=0.47\textwidth]{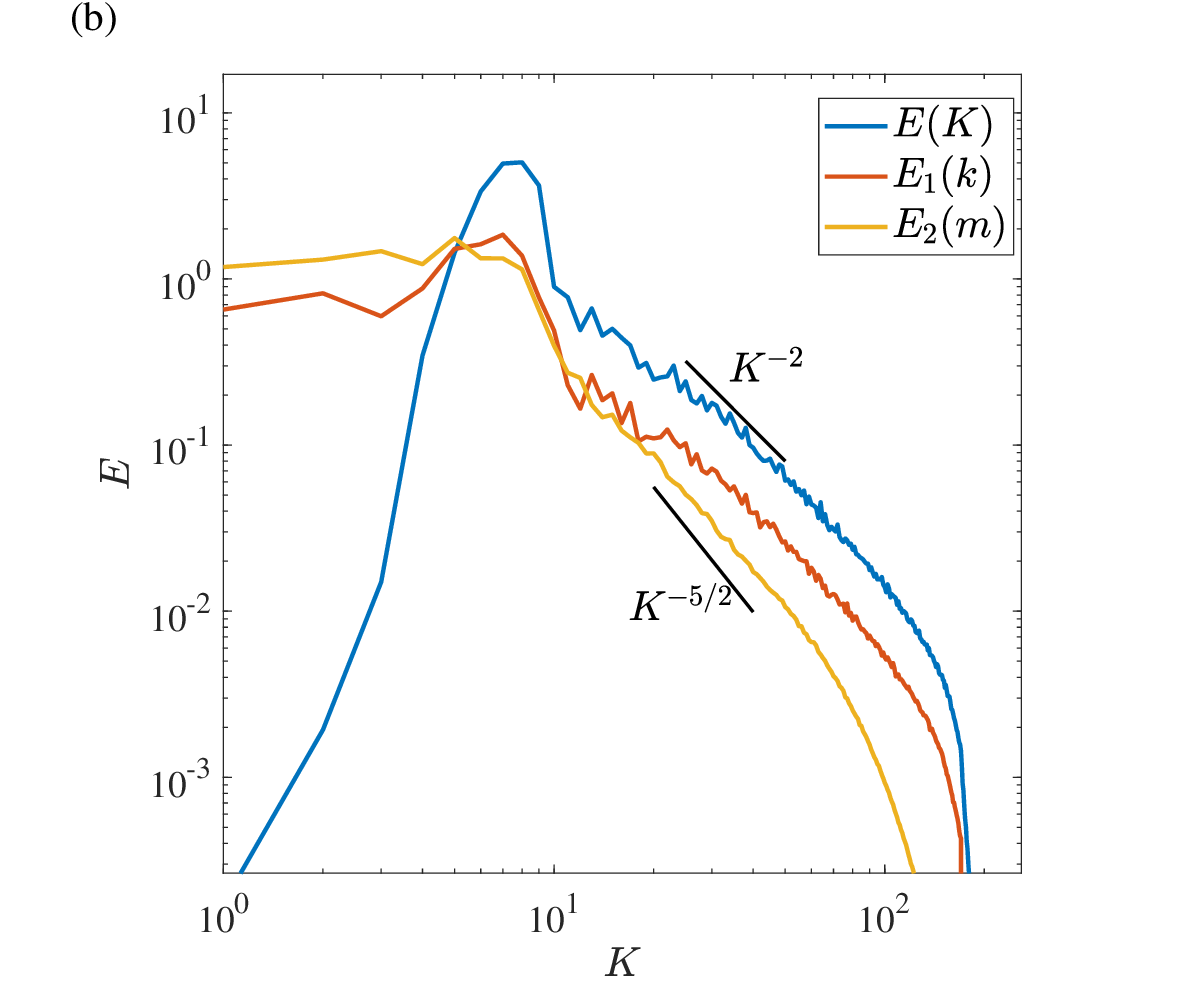}
	\caption{Energe spectra along different orientations. $E_1(k)=\sum_{m}E(k,m)$ along $k$ axis, $E_2(m)=\sum_{k}E(k,m)$ along $m$ axis. (a) The simulation of the original isotropic system (\ref{eq:iso}). (b) The simulation of the anisotropic system (\ref{eq:aniso}).}
	\label{fig:vsspe}
\end{figure}
\begin{figure}
	\centering
	\includegraphics[width=0.47\textwidth]{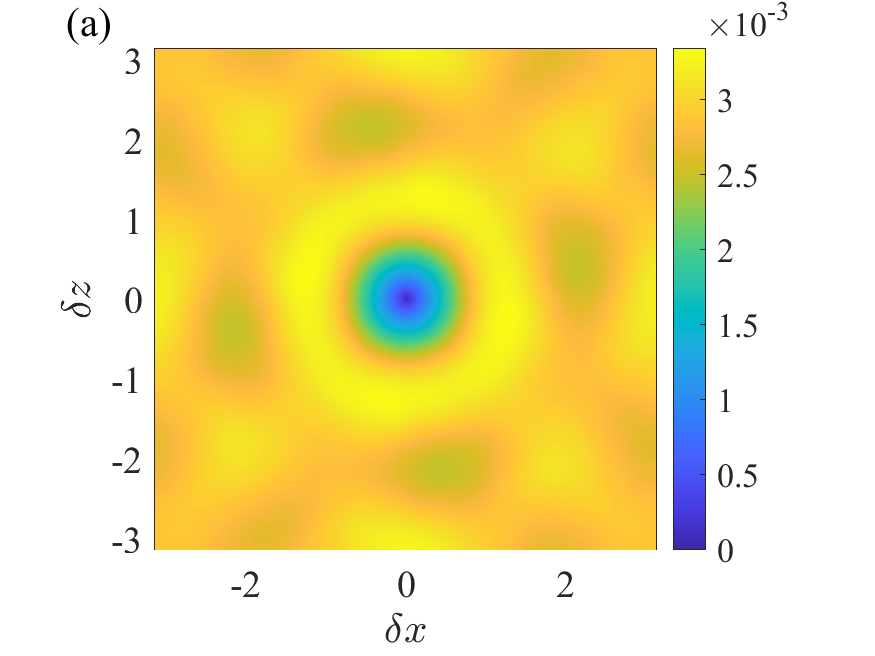}
	\includegraphics[width=0.47\textwidth]{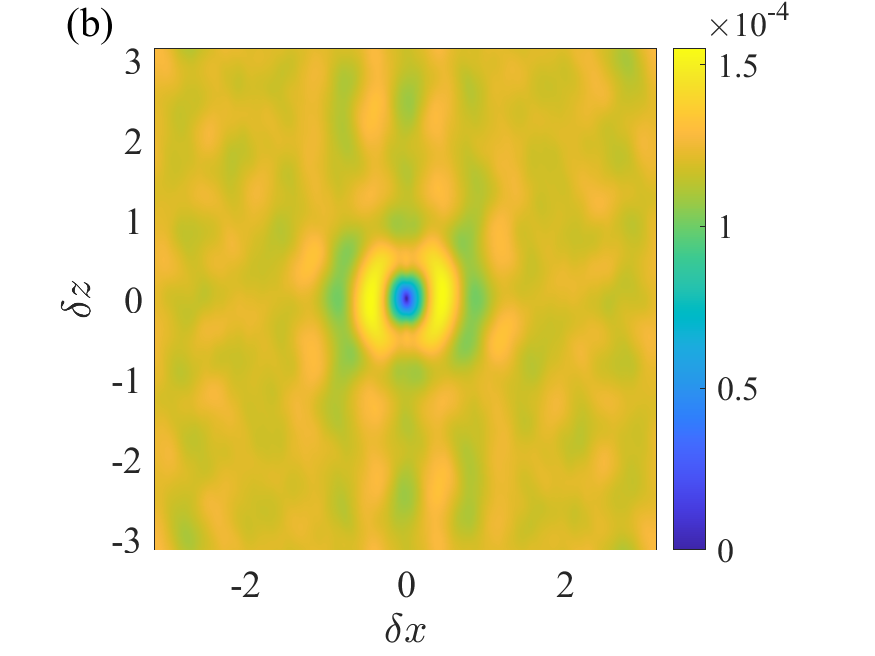}
	\caption{Second-order structure function $\overline{\delta u^2+\delta w^2}$ of the isotropic system (\ref{eq:iso}) (a) and the anisotropic system (\ref{eq:aniso}) (b).}
	\label{fig:vsstf}
\end{figure}

Fig. \ref{fig:vsspe} compares energy spectra along different orientations in the isotropic and anisotropic systems. 
Here, $E_1(k)=\sum_{m}E(k,m)$ and $E_2(m)=\sum_{k}E(k,m)$, where $\sum_{m}$ and $\sum_{k}$ mean the summation over all wavenumber $m$ and $k$, respectively.
In the isotropic system, the curves of $E_1(k)$ and $E_2(m)$ align closely with the same $K^{-2}$ scaling, while in the anisotropic system, the two separate from each other, with $E_1(k)$ scaling as $K^{-2}$ and $E_2(m)$ scaling as $K^{-5/2}$.

\begin{figure}[h]
	\centering
	\begin{subfigure}[b]{0.49\textwidth}
		\centering
		\includegraphics[width=0.8\textwidth]{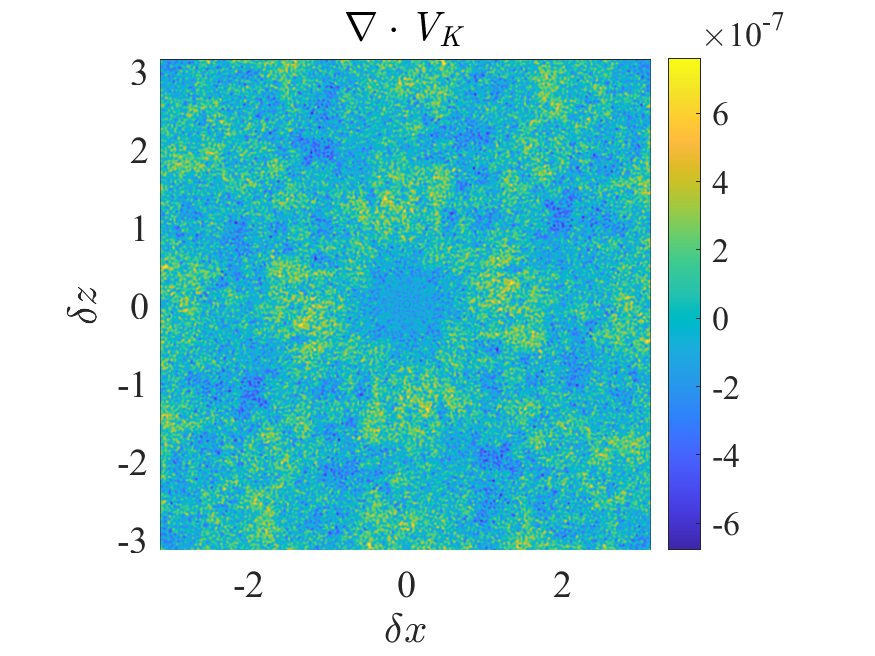}
	\end{subfigure}
	\hfill
	\begin{subfigure}[b]{0.49\textwidth}
		\centering
		\includegraphics[width=0.8\textwidth]{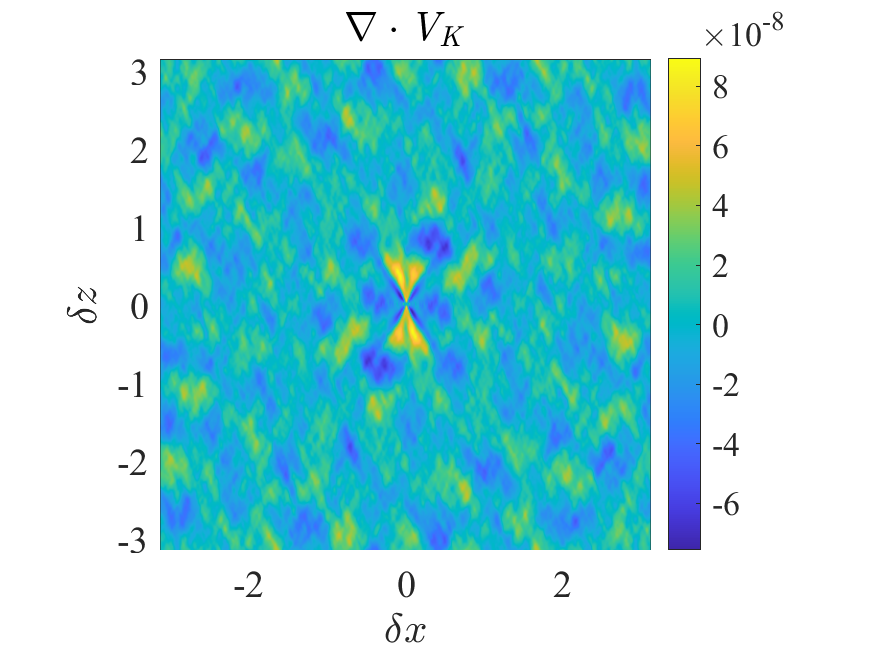}
	\end{subfigure}
	\begin{subfigure}[b]{0.49\textwidth}
		\centering
		\includegraphics[width=0.8\textwidth]{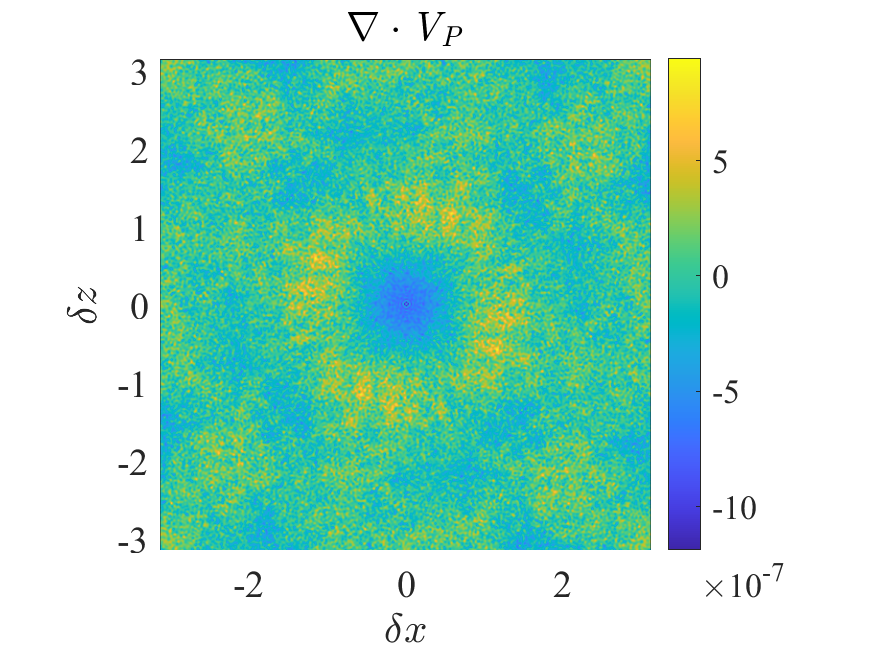}
	\end{subfigure}
	\hfill
	\begin{subfigure}[b]{0.49\textwidth}
		\centering
		\includegraphics[width=0.8\textwidth]{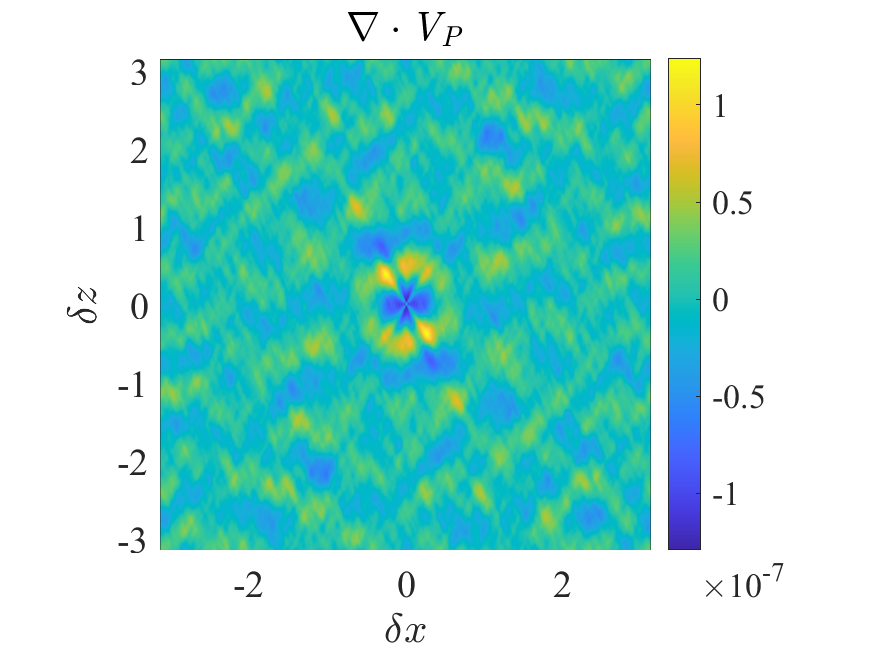}
	\end{subfigure}
	\begin{subfigure}[b]{0.49\textwidth}
		\centering
		\includegraphics[width=0.8\textwidth]{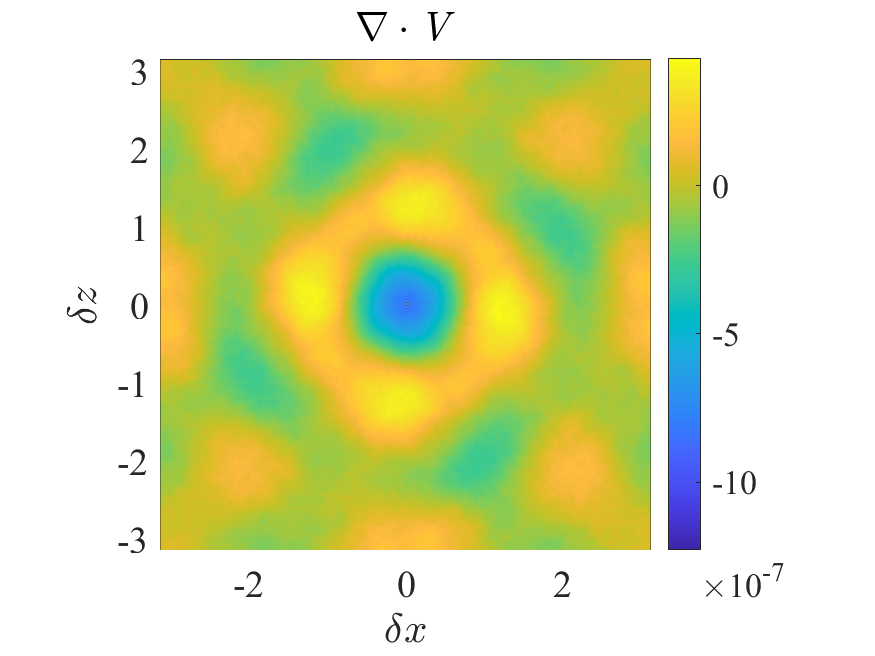}
	\end{subfigure}
	\hfill
	\begin{subfigure}[b]{0.49\textwidth}
		\centering
		\includegraphics[width=0.8\textwidth]{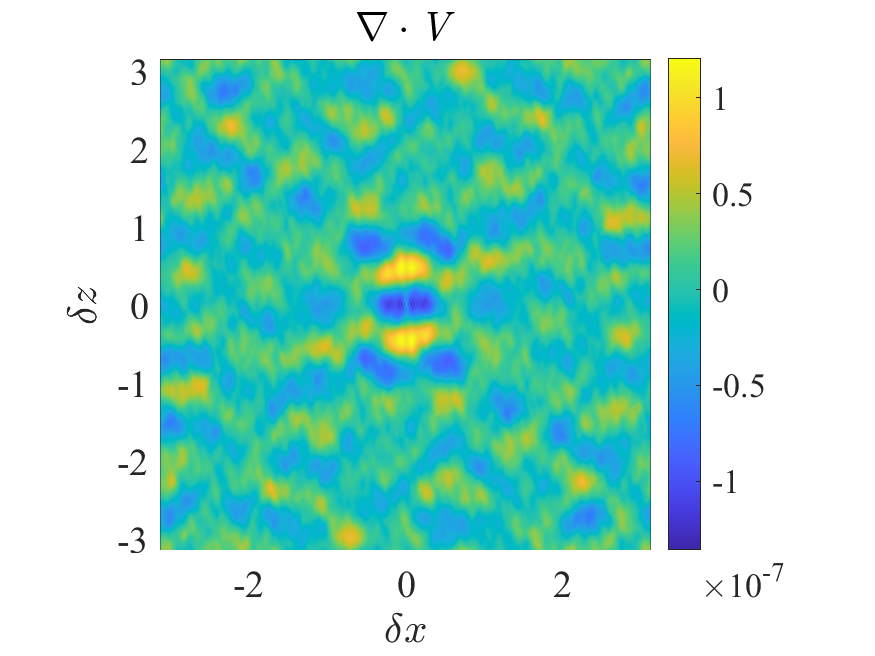}
	\end{subfigure}
	\caption{Third-order structure functions $\nabla\cdot V_K=\nabla\cdot\overline{\delta \bs{u} \br{\delta u^2+\delta w^2}}$, $\nabla\cdot V_P=\nabla\cdot\overline{\delta \bs{u} \br{\delta \theta^2+\delta v^2}}$ and $\nabla\cdot V=\nabla\cdot V_K+\nabla\cdot V_P$ of the isotropic system (\ref{eq:iso}) (left column) and the anisotropic system (\ref{eq:aniso}) (right column).}
	\label{fig:vsstf3}
\end{figure}

We examine second-order structure functions $\overline{\delta u^2+\delta w^2}$, which measures accumulated energy, and third-order structure functions, e.g., $\nabla\cdot V_K=\nabla\cdot\overline{\delta \bs{u} \br{\delta u^2+\delta w^2}}$, $\nabla\cdot V_P=\nabla\cdot\overline{\delta \bs{u} \br{\delta \theta^2+\delta v^2}}$ and $\nabla\cdot V=\nabla\cdot V_K+\nabla\cdot V_P$, which measure energy flux, to provide further verification of isotropy.
Fig. \ref{fig:vsstf} and \ref{fig:vsstf3} show the fields of second and third-order structure functions of the two systems. 
Note that the square computation domain causes the anisotropy at large scales of the isotropic system, and the flocculent structures or narrow cones are the results of resonant wave interactions at very small $K$ close to the forcing scales \cite[cf.][]{kochurin2022direct}. As the anisotropic system, vertical bands appear in Fig. \ref{fig:vsstf}(b). As discussed above, energy accumulates at large scales around $m=0$, thus, $z$-direction bands appear.

\section{\label{Dis}Summary and discussion}

We study the validity of the isotropic arguments in an artificial isotropic rotating fluid system.
In the wave-dominant scenario, i.e., $\Ro\ll 1$, this system has a single time scale for the linear inertial wave and therefore becomes a perfect example that accords with the arguments by \citet{Zeman1994} and \citet{Zhou1995} to obtain the $K^{-2}$ energy spectrum for rotating turbulence.
We asymptotically derive an amplitude equation for the slow evolution of wave amplitude and obtain the $K^{-2}$ energy spectrum using the strong turbulence argument with a constant downscale energy flux.
The statistical isotropy and the $K^{-2}$ scaling are justified in numerical simulations of both the amplitude equation and the isotropic rotating system.
As a comparison, in an anisotropic 2D rotating system energy spectra have different scalings along different directions.

From wave turbulence theory we can also obtain $K^{-2}$ spectrum in our isotropic system (\ref{eq:iso}) (cf. Appendix \ref{WT}). 
Notably, wave turbulence theory generally handle with resonant interaction of weak dispersive waves. 
However, considering the dispersion relation $\omega=\pm f$ where the frequency is independent of wavenumber in the current isotropic rotating system, the quartic interaction does not rule out any interacting modes like what happens in dispersive wave turbulence.
The dispersion relation also implies a zero group velocity which leads to a strong turbulence argument, which differs from the weak wave turbulence.


This study provides a model for theories that require the assumption of isotropy but in fact deal with anisotropic problems. Many theories that involve dimensional analysis or characteristic scales are in this manner. Additionally, the two-dimensional isotropic model presented in this paper can be extended to three-dimensional cases and they present a three-dimensional numerical discussion methodology for theories that incorporate the assumption of isotropy.
Regarding the realization of this two-dimensional isotropic rotating model, the associated experiments are facile to implement, such as confining the rotating stratified flow between two vertically oriented parallel plates or employing a magnetic field to restrict the flow in the y-direction.


\textbf{Acknowledgment}
The authors thank Gregory Falkovich and Sergey Nazarenko for their valuable discussion on wave turbulence and strong turbulence.
This work has received financial support from the National Natural Science Foundation of China (NSFC) under grant NO. 92052102 and 12272006, and from the Laoshan Laboratory under grant NO. 2022QNLM010201.

\appendix

\section{\label{Order1}$O(Ro)$ terms in simulations}

\subsection{\label{Order1ae}The amplitude equation}
\begin{figure}[h]
	\centering
	\includegraphics[width=0.5\textwidth]{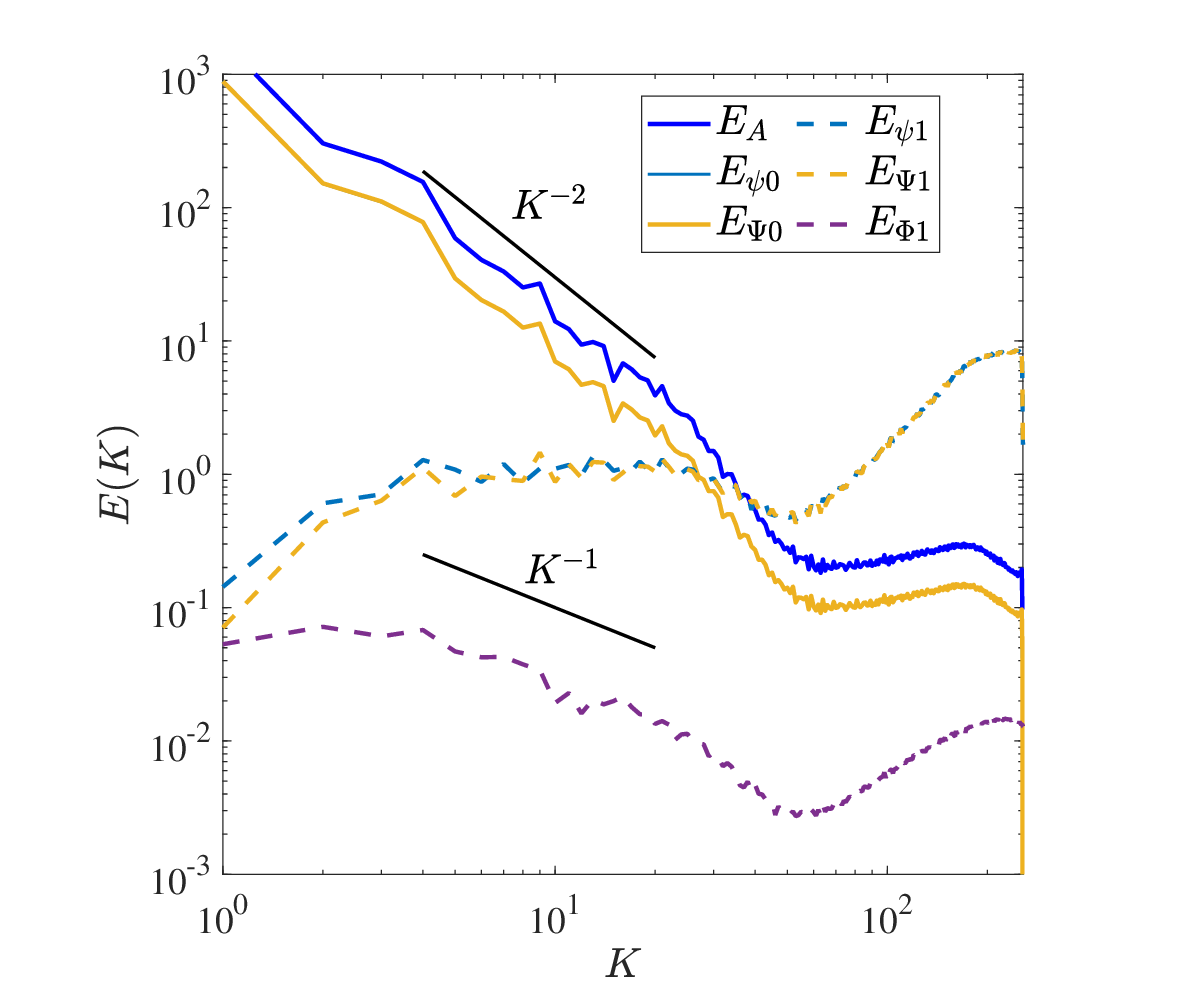}
	\caption{Plot of $E_A$, $E_{\psi0}$, $E_{\Psi0}$, $E_{\psi1}$, $E_{\Psi1}$ and $E_{\Phi1}$. The two solid black lines in Fig. are references for $K^{-2}$ and $K^{-1}$.}
	\label{fig:order01}
\end{figure}

\begin{figure}[h]
	\centering
	\includegraphics[width=0.32\textwidth]{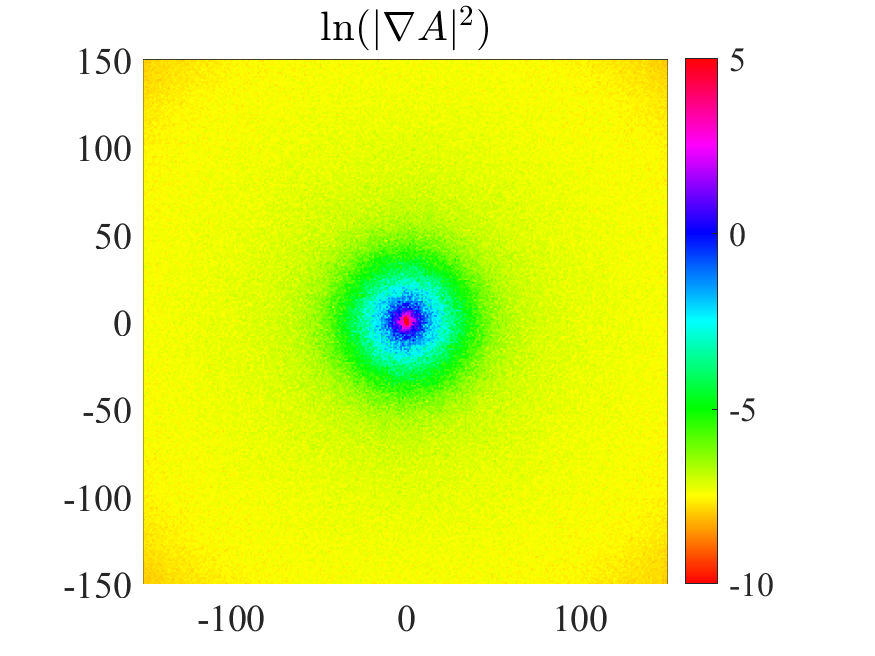}
	\includegraphics[width=0.32\textwidth]{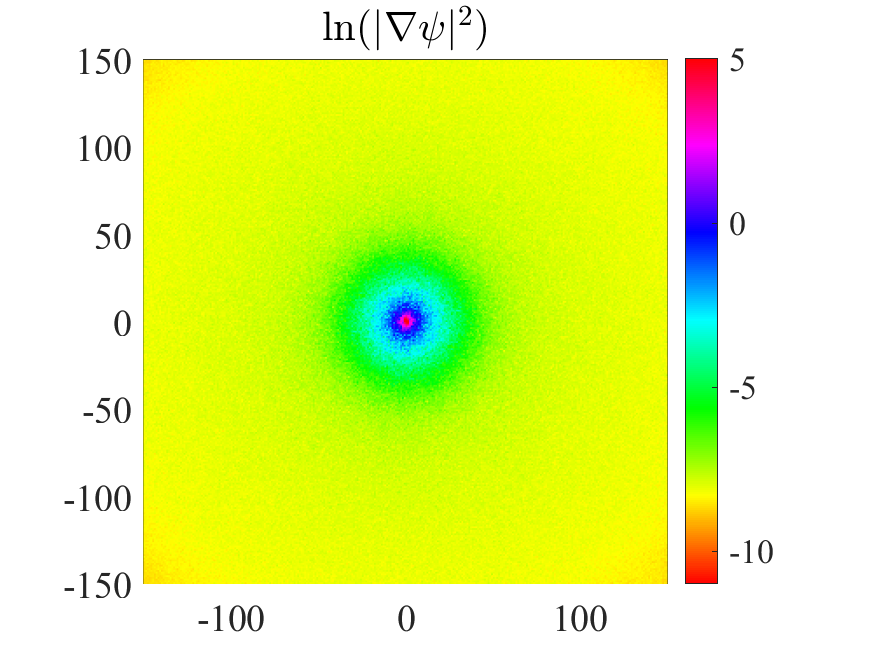}
	\includegraphics[width=0.32\textwidth]{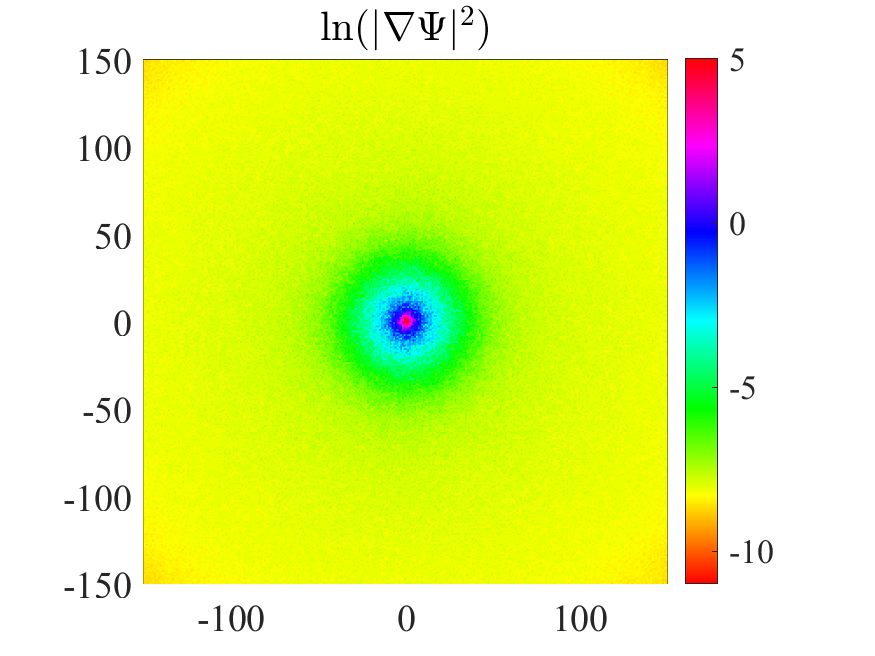}
	\includegraphics[width=0.32\textwidth]{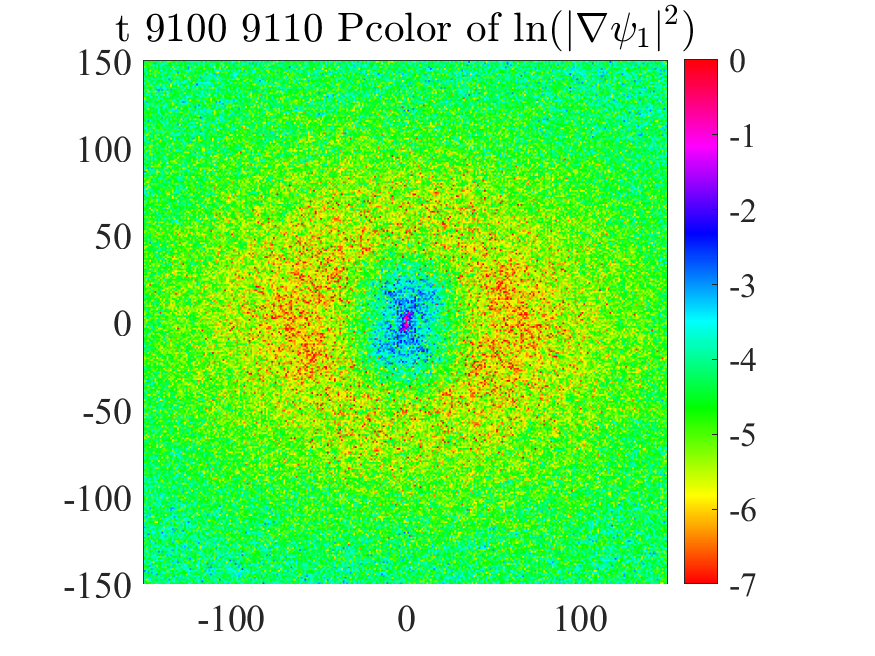}
	\includegraphics[width=0.32\textwidth]{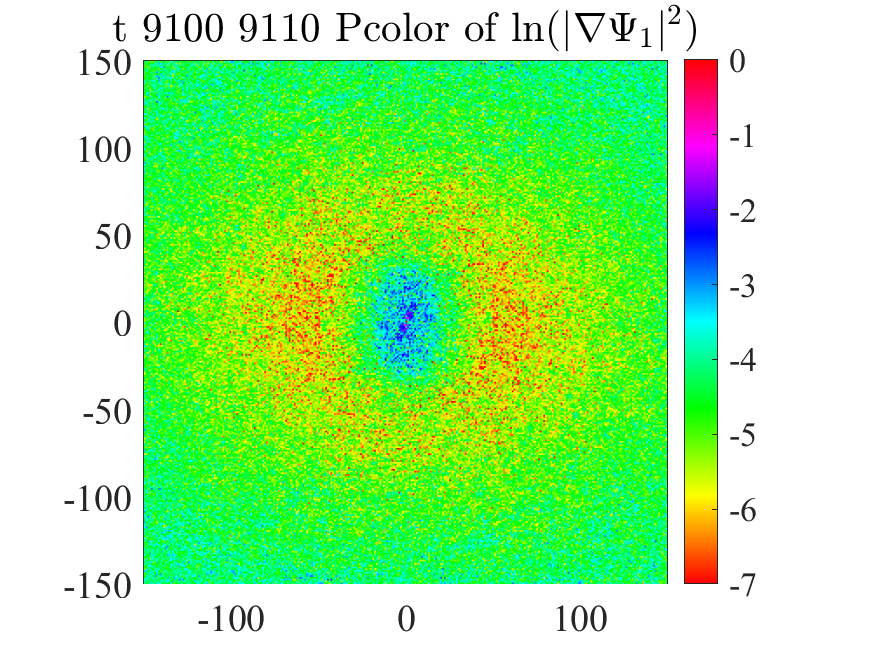}
	\includegraphics[width=0.32\textwidth]{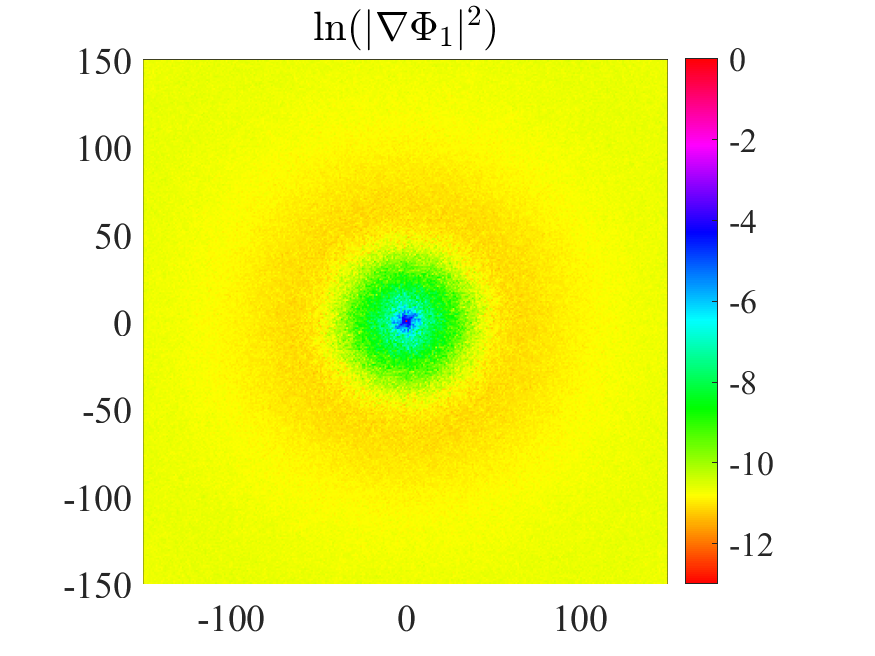}
	\caption{Fields of $\abs{\nabla A}^2$, $\abs{\nabla \psi}^2$, $\abs{\nabla \Psi}^2$, $\abs{\nabla \psi_1}^2$, $\abs{\nabla \Psi_1}^2$, $\abs{\nabla \Phi_1}^2$, averaged over $t\in[7100,7900]$. The horizontal axis is $k$-axis and the vertical axis is $m$-axis. }
	\label{fig:aepcolor}
\end{figure}

Now that the spectrum of $E_A$ scales as $K^{-2}$, which implies $A\sim K^{-3/2}$, we can deduce by dimension analysis from (\ref{eq:order0}) and (\ref{eq:order1}) that the spectra of leading terms $E_{\psi 0}$ and $E_{\Psi 0}$ should scale as $K^{-2}$ and the spectra of order-one terms $E_{\psi1}$, $E_{\Psi{1}}$ and $E_{\Phi1}$ should scale as $K^{-1}$. For example, substituting $A\sim K^{-3/2}$ into equation (\ref{eq:order1psi}) we obtain $\psi_{1}\sim K^{-1}$, then considering $E_{\psi 1} \mdd K \sim \nabla\psi_{1}\cdot\nabla\psi_{1}^*$ we get $E_{\psi 1}\sim K^{-1}$.
We numerically calculate the corresponding spectra of $\psi_0$, $\Psi_0$, $\psi_1$, $\Psi_{1}$ and $\Phi_{1}$ in Fig. \ref{fig:order01}. 
We then plot the fields of time-averaged spectra in Fig. \ref{fig:aepcolor}. 


\subsection{\label{Order1oe}The original euqations}

\begin{figure}[h]
	\centering
	\includegraphics[width=0.5\textwidth]{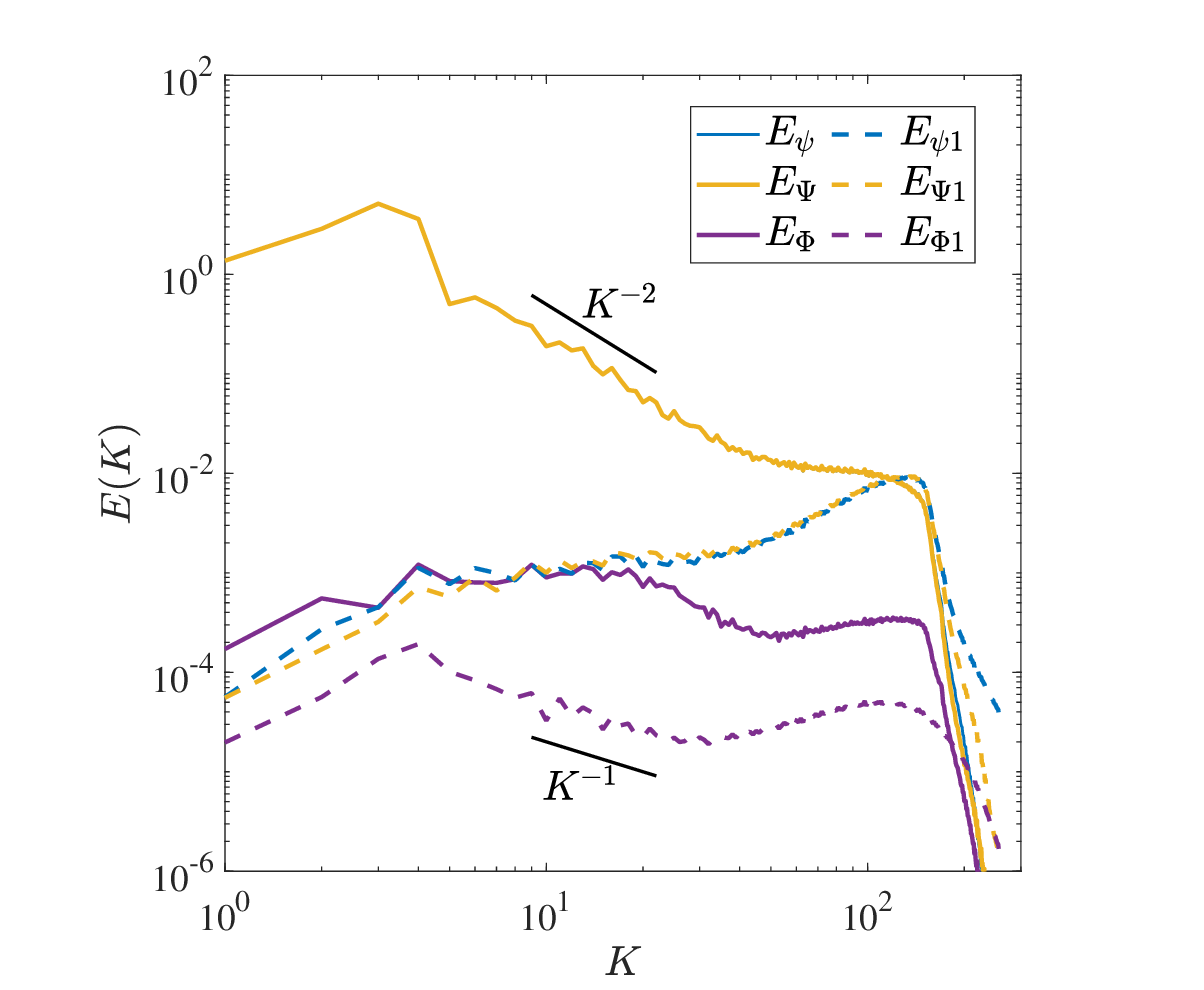}
	\caption{Energy spectra $E_A$, $E_{\psi0}$, $E_{\Psi0}$, $E_{\psi1}$, $E_{\Psi1}$ and $E_{\Phi1}$. The two solid black lines are $K^{-2}$ and $K^{-1}$ as references.}
	\label{fig:oeorder01spe}
\end{figure}

From (\ref{eq:order0}) and (\ref{eq:order1}) we calculate the spectra of $O(1)$ terms and show them in Fig. \ref{fig:oeorder01spe}, where we approximate $\psi_0$ as $\psi$ and $\Psi_0$ as $\Psi$. The results are similar to Fig. \ref{fig:order01} of the simulation of amplitude equation.

\begin{figure}[h]
	\centering
	\includegraphics[width=0.32\textwidth]{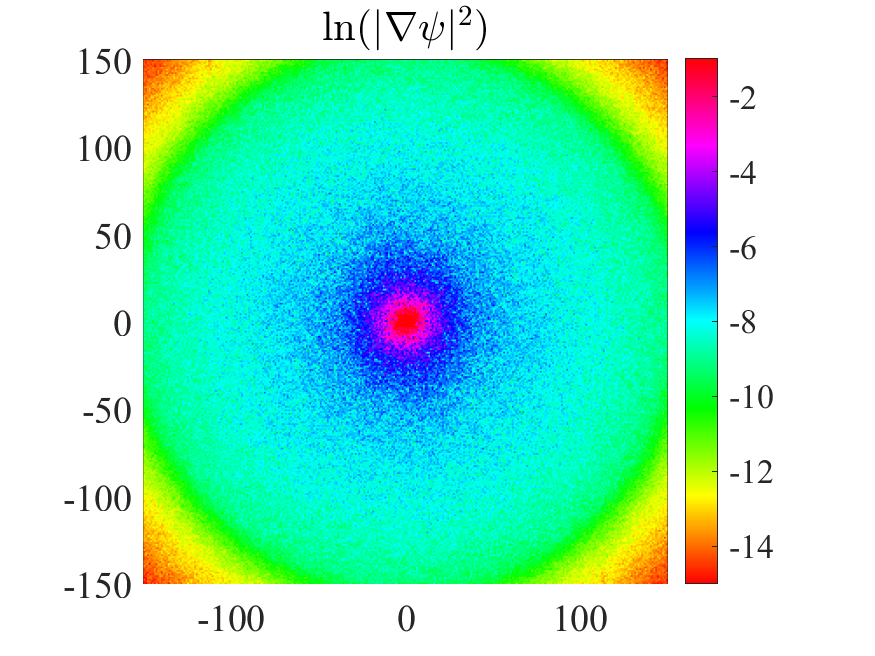}
	\includegraphics[width=0.32\textwidth]{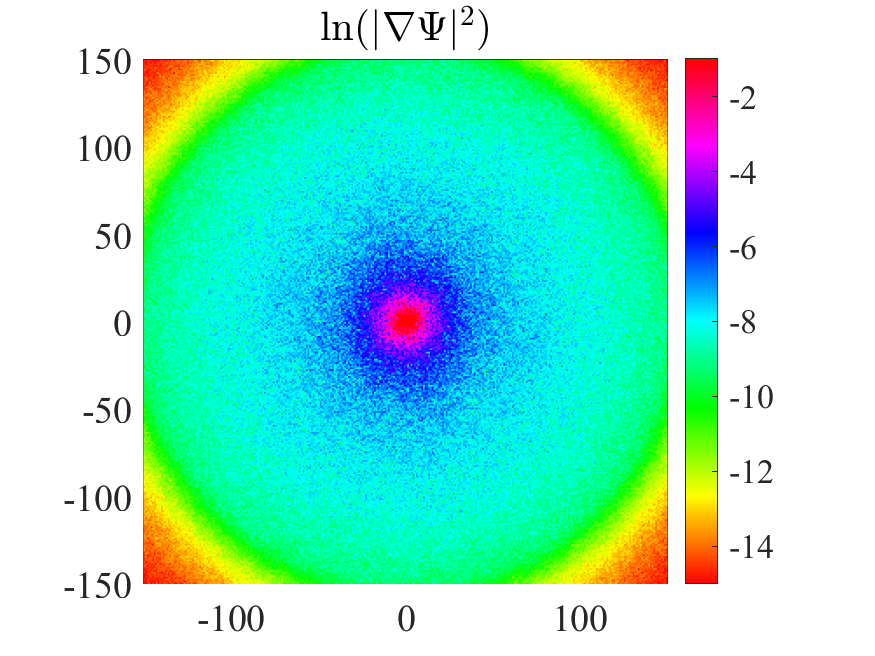}
	\includegraphics[width=0.32\textwidth]{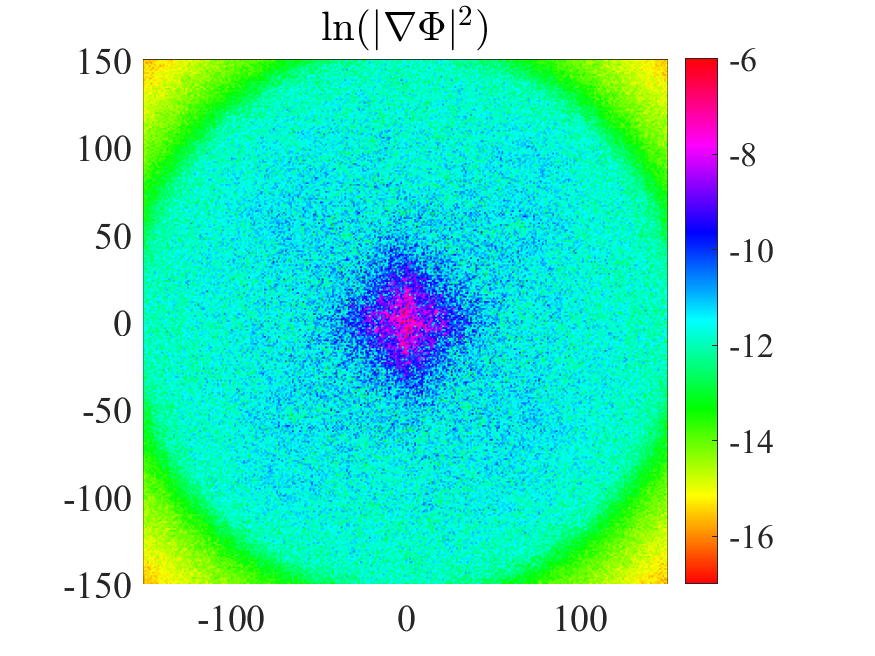}
	\includegraphics[width=0.32\textwidth]{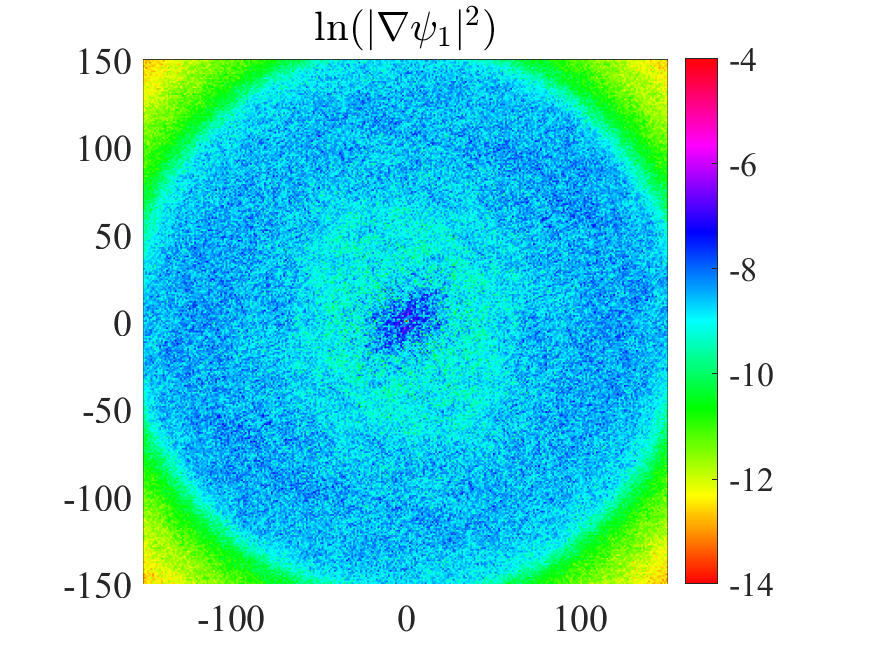}
	\includegraphics[width=0.32\textwidth]{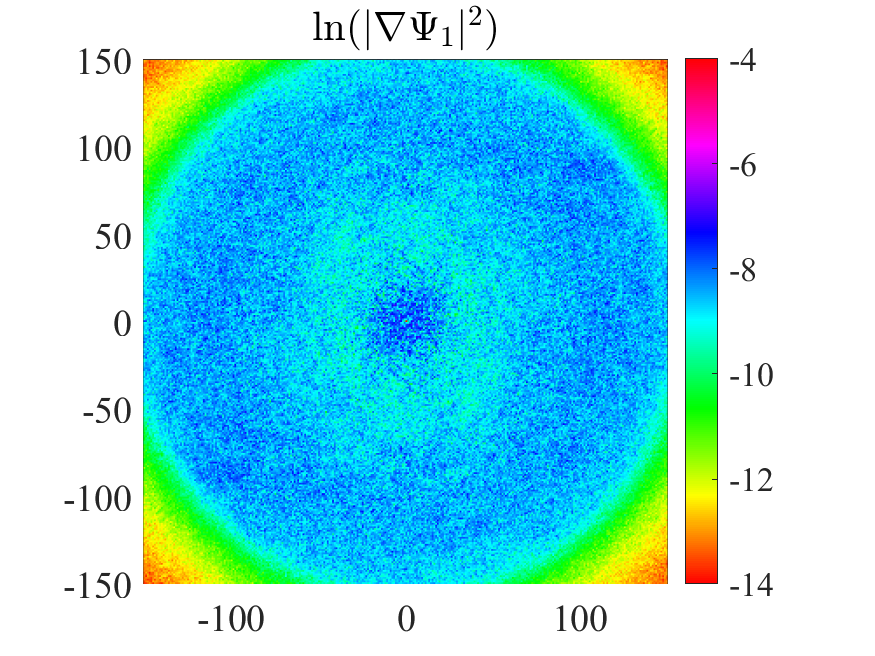}
	\includegraphics[width=0.32\textwidth]{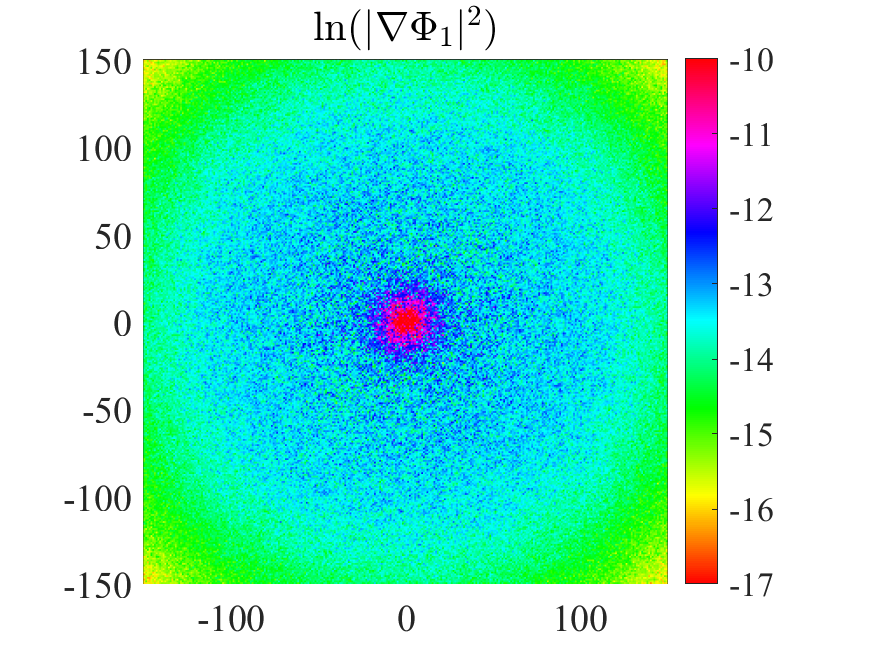}
	\caption{Fields of $\abs{\nabla \psi}^2$, $\abs{\nabla \Psi}^2$, $\abs{\nabla \Phi}^2$, $\abs{\nabla \psi_1}^2$, $\abs{\nabla \Psi_1}^2$, $\abs{\nabla \Phi_1}^2$. The horizontal axis is $k$-axis and the vertical axis is $m$-axis. }
	\label{fig:oepcolor}
\end{figure}

Since the $O(\Ro)$ terms are terms with $\omega=2$ and $\omega=0$ (cf. (\ref{eq:order1})), the behaviours of spectra of $\psi_{1}$ and $\Psi_{1}$ in Fig. \ref{fig:oeorder01spe} and $E_{\text{tot}}|_{w=2.0}$ in Fig. \ref{fig:seperatespe}(b) are consistent.
However, $E_{\psi1}$ and $E_{\Psi1}$ (and $E_{\text{tot}}|_{w=2.0}$) should scale as $K^{-1}$ (cf. (\ref{eq:order1})), rather than $K^0$ in Fig. \ref{fig:seperatespe}(b), Fig. \ref{fig:oeorder01spe} and Fig. \ref{fig:order01}. 
We guess that the phase distribution of the amplitude is not uniform enough to support our dimension analysis for order-one terms.
Also, we plot the fields of time-averaged spectra in Fig. \ref{fig:oepcolor}. The anisotropy of $\abs{\nabla \psi_1}^2$ and $\abs{\nabla \Psi_1}^2$ show that $\psi_{1}$ and $\Psi_{1}$ have nonuniform phase distributions, implying that the phase distribution of the amplitude is not uniform enough, which supports our guess.
In addition, the field of $\abs{\nabla \Phi}^2$ is also anisotropic but seems a square, which could be caused by the square computation domain.

\section{\label{WT} Scalings of energy spectrum Obtained by Wave Turbulence Theory}
Dimensional analysis in WT theory (\cite{Nazarenko2011wave}) shows that for general $d$-dimensional wave systems, the energy cascade spectrum has the scaling
\begin{equation*}
	E_K^{(\textrm{1D})}\sim \lambda^x \epsilon^{1/(N-1)} K^y,
\end{equation*}
where 
\begin{equation*}
	x=2-\frac{3}{N-1},\qquad \quad y=d-6+2\alpha+\frac{5-d-3\alpha}{N-1},
\end{equation*}
$N$ is the minimal number of resonant waves, $\epsilon$ is energy injection rate, and $\lambda$ and $\alpha$ come from dispersion relation $\omega=\lambda K^{\alpha}$. 

In classic 3D anisotropic rotating turbulence, the dispersion relation is 
$$\omega=f\frac{k_{\parallel}}{K}$$ ($\parallel$ refers to the rotation axis). If we consider an isotropic assumption that $k_{\parallel}\sim k_{\perp}$, we can take $\lambda\sim f$ and $\alpha=0$. With $d=3$ and $N=3$, we get $$E_K^{(\textrm{1D})}\sim f^{1/2} \epsilon^{1/2} K^{-2},$$ which is consistent with \citet{Zeman1994} and \citet{Zhou1995} in 3D condition.

In our system where the dispersion relation is $\omega=\pm f$, with $d=2$, $N=4$, $\lambda=f$, $\alpha=0$, we obtain
\begin{equation*}
	E_K^{(\textrm{1D})}\sim f \epsilon^{1/3} { K^{-3}}.
\end{equation*}
The scaling is not $K^{-2}$ but $K^{-3}$. It is caused by different definitions of $E(K)$ in WT theory and in our derivation
 \begin{subequations}
	\begin{align*}
		\text{WT: }\sbr{\int E_K^{(\textrm{1D})} \mdd K}&=\frac{\sbr{\frac{1}{2}m u^2}}{\sbr{l^d}}=\sbr{l^{3-d} u^2}\\
		\text{Our: }\sbr{\int E(K) \mdd K}&=\sbr{ u^2}
	\end{align*}
\end{subequations}
When $d=2$,
\begin{equation*}
	\sbr{E(K)}=\sbr{E_K^{(\textrm{1D})}K}\sim K^{-2}.
\end{equation*}
Thus we get $K^{-2}$ scaling, which is consistent with \citet{Zeman1994} and \citet{Zhou1995} in 2D condition.

%
%
%


\bibliography{WT3.bib}

\end{document}
%